\documentclass[11pt]{article}

\usepackage[margin=1in]{geometry}
\usepackage{amsmath,amssymb,amsthm}
\usepackage{booktabs}
\usepackage{graphicx}
\usepackage[caption=false]{subfig}

\usepackage[natbibapa,nodoi]{apacite}
\usepackage{bm}
\usepackage{algorithm}
\usepackage{algpseudocode}
\usepackage{float}
\usepackage{siunitx}
\usepackage{hyperref}

\theoremstyle{plain}

\theoremstyle{definition}

\theoremstyle{remark}

\renewcommand{\d}{\,\mathrm{d}}

\begin{document}

\title{Using Variational Inference to Improve the Efficiency of MCMC Algorithms}

\author{Pingping Yin\textsuperscript{a} and Xiyun Jiao\textsuperscript{a}\thanks{CONTACT Xiyun Jiao. Email: jiaoxy@sustech.edu.cn}\\[0.3em]
\small\textsuperscript{a}Department of Statistics and Data Science,\\
\small Southern University of Science and Technology, Shenzhen, China}
\date{}

\maketitle

\begin{abstract}
    Bayesian statistics makes inference based on Bayes' theorem, but the posterior distribution of unknown parameters is typically analytically intractable. To estimate the posterior, two widely used numerical approximation methods are Markov Chain Monte Carlo (MCMC) and variational inference (VI). MCMC methods produce asymptotically exact samples but are computationally intensive, while VI methods are faster and more scalable but may lack accuracy. This paper proposes combining MCMC and VI to construct algorithms that leverage the strengths of both. The first proposed algorithm uses Gaussian variational inference (GVI) with various covariance structures to derive a linear transformation matrix for Hamiltonian Monte Carlo (HMC). This method improves the efficiency of HMC, particularly in high-dimensional and complex target distributions. The second algorithm combines a VI-based generative model, the variational auto-encoder (VAE), with the Metropolis-Hastings (MH) sampler. The resulting VAE-MH sampler is efficient and effectively traverses the parameter space, outperforming standard MCMC methods in identifying all modes of multi-modal distributions. Code implementations are available at \href{https://github.com/YinPingping111/HMC-GVI}{HMC-GVI} and \href{https://github.com/YinPingping111/VAE-MH-Sampler}{VAE-MH-Sampler}.
\end{abstract}

\noindent\textbf{Keywords:}
Bayesian inference; Efficiency; Gaussian Variational Inference; Markov Chain Monte Carlo; Variational Auto--Encoder

\section{Introduction}
\label{sec1}
Bayesian statistics makes inference based on the posterior distribution of the unknown parameters. Suppose $\bm{x} = (\bm{x}_1, \dots, \bm{x}_n)$ is the data and $\bm{\theta}\in \Theta \subset \mathbb{R}^d$ is the vector of unknown parameters. We specify the likelihood function as $p(\bm{x}|\bm{\theta})$ and the prior distribution for $\bm{\theta}$ as $p(\bm{\theta})$. Then the posterior distribution of $\bm{\theta}$ is
$$
\pi(\bm{\theta}):=p(\bm{\theta}|\bm{x})=\frac{p(\bm{\theta})p(\bm{x}|\bm{\theta})}{p(\bm{x})}=
\frac{\tilde{\pi}(\bm{\theta})}{\int_{\Theta}\tilde{\pi}(\bm{u})\d\bm{u}}
$$
where $\tilde{\pi}(\bm{\theta})=p(\bm{\theta})p(\bm{x}|\bm{\theta})$ is the unnormalized posterior density,  $p(\bm{x})\equiv \int p(\bm{\theta})p(\bm{x}|\bm{\theta})
\d\bm{\theta}$ is the \emph{marginal likelihood} (or \emph{evidence}). Throughout this paper, we denote the target density by $\pi$, its unnormalized version by $\tilde{\pi}$, and the log unnormalized target density by $\ell(\bm{\theta})=\log \tilde{\pi}(\bm{\theta})$. Because $p(\bm{x})$ is often analytically intractable, the posterior distribution $p(\bm{\theta}|\bm{x})$ is not in closed form. There are two widely used approaches to approximating $p(\bm{\theta}|\bm{x})$, which are Markov Chain Monte Carlo (MCMC) and variational inference (VI), each having its own strengths and weaknesses. 

MCMC is a class of sampling-based methods, constructing a Markov chain $\{\bm{\theta}^{(t)}, t=1, 2, \dots\}$ for which the stationary distribution is the target posterior $\pi(\bm{\theta})$. Metropolis-Hastings (MH) \citep{J_JCP_metropolis1953equation,J_BIOMET_hastings1970monte} is one of the most widely used MCMC methods. MH produces samples from the target distribution by first generating candidates from a proposal distribution and accepting or rejecting with a probability. The most common choice of the proposal is random walk (RW), which is a Gaussian or uniform distribution centered around the current state $\bm{\theta}^{(t)}$. The MH algorithm with the RW proposal is easy to implement, but inefficient for high-dimensional or multi-modal distributions due to its local and random nature. Furthermore, scale heterogeneity and high correlations among the parameters may largely exacerbate the issue.

Many new proposals have been developed to mitigate the problems associated with RW. Metropolis Adjusted Langevin algorithm (MALA) and Hamiltonian Monte Carlo (HMC) both use the gradient information of the target distribution in their proposals to generate candidates closer to high-density regions. MALA and HMC are much more robust to dimensionality compared with RW. Asymptotically, as the data size $n\to \infty$, the efficiencies of MALA and HMC decay with the dimension of $\bm{\theta}$ in the order of $d^{-1/3}$ and $d^{-1/4}$ respectively \citep{J_JRSSSB_roberts1998optimal, J_BERNOULLI_beskos2013optimal}, while the efficiency of RW scales by $d^{-1}$ \citep{J_AOAP_gelman1997weak}. 

MALA and HMC avoid proposing randomly, but they explore the parameter space locally, as RW does, and are therefore not effective in sampling from multi-modal distributions either \citep{C_NeurIPS_samsonov2022localglobal,C_AAAI_lan2014wormhole}. Recent efforts to improve the efficiency of sampling from multi-modal distributions focus on combining global and local proposals. For multi-modal targets, an effective sampler usually needs to locate the important modes, move between them with global proposals, and explore locally within each mode. Recently, generative models have been widely used as global proposals to assist local MCMC algorithms for sampling from multi-modal posteriors, due to their strong ability to learn complex probability distributions underlying a sample. For example, \citet{C_ICMLW_gabrie2021efficient} assisted MALA by normalizing flow (NF), and \citet{J_CPC_hunt-smith2024accelerating} used the diffusion model to improve the mixing of the MH algorithm with the RW proposal. However, these generative-model proposals still rely on the training samples or initial states to cover the important modes. When the modes are far apart, the important mode locations often need to be available \emph{a priori}, and the mode-finding step is not provided as a complete component of the sampler.

A common way to handle the heterogeneity of scales and high correlations among parameters is applying a matrix transformation to $\bm{\theta}$ so that the transformed parameters are approximately independent with unit standard deviation. One popular choice for such a matrix is the Cholesky decomposition of the covariance matrix of $\bm{\theta}$ estimated using the burn-in. \citet{J_CSTAT_haario1999adaptive, J_BERNOULLI_haario2001adaptive} introduced two versions of adaptive RW Metropolis algorithms, both updating the covariance matrix of the Gaussian proposal continuously to adjust to the target distribution. While the adaptive proposal (AP) of \citet{J_CSTAT_haario1999adaptive} updates the covariance matrix using a fixed number of previous iterations, the adaptive Metropolis (AM) of \citet{J_BERNOULLI_haario2001adaptive} uses all previous iterations. The theoretical arguments in \citet{J_BERNOULLI_haario2001adaptive} guarantee the good properties of AM and the numerical results there show its robust performance in practice compared to the standard RW and AP. However, for high-dimensional targets, the adaptation of AM becomes slower and its sensitivity to a bad choice of initial covariance increases.

VI, on the other hand, is an optimization-based approach which searches for the best approximating distribution $q^*(\bm{\theta})$ that minimizes the Kullback-Leibler (KL) distance to the target posterior $\pi(\bm{\theta})$ among a pre-specified tractable distribution family $\mathcal{Q}$ \citep{bishop2006pattern}. Compared to MCMC, VI is much more efficient in computation and scalable with dimension. However, the performance of VI largely depends on the specification of the variational family $\mathcal{Q}$. Mean-field VI assumes each distribution in $\mathcal{Q}$ to have the form $q(\bm{\theta})=\prod_{j=1}^d q_j(\bm{\theta}_j)$, i.e., the components of $\bm{\theta}$ are mutually independent and each governed by a distinct factor in the variational density \citep{J_JASA_blei2017variational}. Despite its advantage in computation, mean-field VI ignores the dependence between parameters, which may lead to significant bias. To address this limitation, many authors have proposed more sophisticated variational densities which reflect the posterior dependence structure to varying degrees \citep{gershman2012tutorial, salimans2013fixed}. Normalizing flows enrich variational families by transforming a simple base density through a sequence of invertible maps, with the change-of-variables formula keeping the resulting density evaluation tractable \citep{C_ICML_rezende2015variational}. Gaussian variational inference (GVI) instead takes the variational distribution to be Gaussian, for example $q_{\bm{\lambda}}(\bm{\theta})=\mathcal N(\bm{\mu},\bm{\Sigma})$, and uses structured covariance or precision-matrix parameterizations to balance posterior dependence representation and computational scalability \citep{arXiv_tran2021practical,J_JCGS_ong2018gaussian,J_SAC_tan2018gaussian}. Variational auto-encoders (VAEs) apply amortized VI to deep latent-variable models by introducing an encoder $q_{\bm{\phi}}(\bm{z}\mid\bm{x})$ and a decoder $p_{\bm{\eta}}(\bm{x}\mid\bm{z})$, whose parameters are learned jointly by maximizing the evidence lower bound \citep{C_ICLR_kingma2014autoencoding}. Variants based on normalizing flows, mixture latent structures, or heavy-tailed decoders further improve the flexibility of VAEs for complex distributions \citep{C_ICML_rezende2015variational,C_ECAI_lavda2020improving,C_IPTA_ye2020mixtures,C_IJCAI_takahashi2018student}.

Given their complementary strengths and weaknesses, combining MCMC and VI can potentially yield more efficient algorithms. However, previous attempts to use VI approximations as proposal distributions for MH samplers have faced scalability and mixing efficiency issues. In contrast, VI can be used not only as a direct approximation to the posterior, but also as a source of structural information for MCMC. GVI gives a fast estimate of the location and covariance structure of the target distribution, which is valuable for constructing linear transformations or preconditioning matrices in gradient-based samplers. VAE-type generative models can generate global candidate points from a learned approximation to the target distribution, which is useful for moving between distant modes.

In this paper, we propose two novel approaches that effectively combine VI and MCMC.
\begin{enumerate}
  \item \textbf{Linear Transformation in HMC algorithm via GVI}: GVI with various covariance structures is used to derive a preconditioning matrix for HMC, specifically, we consider GVI with Cholesky decomposed covariance \citep{arXiv_tran2021practical}, GVI with factor decomposed covariance \citep{J_JCGS_ong2018gaussian} and GVI with sparse precision matrices \citep{J_SAC_tan2018gaussian}. This approach addresses the challenge of computing linear transformation matrices for high-dimensional models \citep{J_JRSSSB_girolami2011riemann}. Tests show that HMC with GVI-derived linear transformation matrices outperforms standard MCMC algorithms, especially as the dimension and complexity of the target distribution increase.

   \item \textbf{VAE-MH Sampler}: We develop a generative-proposal MH sampler \citep{J_BIOMET_hastings1970monte} for multi-modal targets. The method first uses CrowdingDE \citep{C_CEC_thomsen2004multimodal} to identify high-probability regions and construct an initial training set around the estimated modes. It then trains a VAE-type generative model \citep{C_ICLR_kingma2014autoencoding} to produce global proposal candidates, applies an explicit KDE-based MH correction, and combines the global kernel with a local random-walk kernel for within-mode exploration. The implementation includes standard VAEs, flow-augmented VAEs, component-conditioned GMM-VAEs, and Student-$t$ decoder variants for heavy-tailed targets. Tests on multi-dimensional mixture Gaussian and mixture $t$ distributions with distant modes demonstrate that VAE-MH can recover the important modes and improve cross-mode mixing, whereas MH often gets trapped in local modes.
\end{enumerate}

The paper is organized as follows. Section 2 provides an overview of MCMC and VI algorithms. Section 3 introduces two novel approaches that effectively combine VI and MCMC: (1) linear transformation in HMC via GVI, and (2) the VAE-MH sampler that combines VAEs with Metropolis-Hastings sampling. Section 4 presents numerical studies evaluating both methods across various challenging distributions. The paper concludes with a discussion in Section 5.

\section{Preliminaries}
\label{section2}
\subsection{Hamiltonian Monte Carlo algorithm}
\label{sec2.1}
HMC avoids the diffusive behavior of RW by augmenting the target distribution with auxiliary momentum variables $\bm{\psi}$ and using discretized Hamiltonian dynamics to generate distant proposals \citep{BS_neal2011mcmc}. Given a positive definite mass matrix $\mathbf M$, the momentum is drawn independently as $\bm{\psi}\sim \mathcal N(\mathbf 0,\mathbf M)$. The Hamiltonian is $H(\bm{\theta},\bm{\psi})=U(\bm{\theta})+K(\bm{\psi})$, where $U(\bm{\theta})=-\ell(\bm{\theta})$ and $K(\bm{\psi})=\frac12\bm{\psi}^{\top}\mathbf M^{-1}\bm{\psi}$. The corresponding joint density is proportional to
$$
\begin{aligned}
\exp\{-H(\bm{\theta},\bm{\psi})\}
&= \tilde{\pi}(\bm{\theta})\exp\{-K(\bm{\psi})\},
\end{aligned}
$$
so that the marginal density of $\bm{\theta}$ is the desired target $\pi$.

The Hamiltonian dynamics satisfy
$$
\frac{d\bm{\theta}}{dt}=\mathbf M^{-1}\bm{\psi},
\qquad
\frac{d\bm{\psi}}{dt}=\nabla_{\bm{\theta}}\ell(\bm{\theta}).
$$
Since these differential equations are rarely solvable in closed form, HMC uses a reversible and volume-preserving leapfrog integrator. Starting from $(\bm{\theta}_0,\bm{\psi}_0)$, where $\bm{\theta}_0$ is the current state and $\bm{\psi}_0$ is the newly drawn momentum, one leapfrog step from $(\bm{\theta}_l,\bm{\psi}_l)$ to $(\bm{\theta}_{l+1},\bm{\psi}_{l+1})$ with step size $\epsilon$ is
$$
\begin{aligned}
    \bm{\psi}_{l+\frac12}
    &=
    \bm{\psi}_l
    +
    \frac{\epsilon}{2}
    \nabla_{\bm{\theta}}\ell(\bm{\theta}_l),\\
    \bm{\theta}_{l+1}
    &=
    \bm{\theta}_l
    +
    \epsilon \mathbf M^{-1}\bm{\psi}_{l+\frac12},\\
    \bm{\psi}_{l+1}
    &=
    \bm{\psi}_{l+\frac12}
    +
    \frac{\epsilon}{2}
    \nabla_{\bm{\theta}}\ell(\bm{\theta}_{l+1}).
\end{aligned}
$$
After $L$ leapfrog steps, we set $(\bm{\theta}^{\star},\bm{\psi}^{\star})=(\bm{\theta}_L,-\bm{\psi}_L)$, where the momentum flip is used for reversibility and does not change the kinetic energy. The proposal is accepted with probability
$$
    \alpha_{\mathrm{HMC}}
    =
    \min\left\{
    1,\,
    \exp\left[
        -H(\bm{\theta}^{\star},\bm{\psi}^{\star})
        +
        H(\bm{\theta}_0,\bm{\psi}_0)
        \right]
    \right\}.
$$
The exact Hamiltonian flow preserves both the Hamiltonian energy and phase-space volume. In practice, the leapfrog integrator is reversible and volume-preserving but introduces discretization error in the Hamiltonian. The final Metropolis--Hastings correction corrects for this discretization error and guarantees that the resulting Markov chain has the desired target distribution as its invariant distribution. Compared with RW, HMC uses gradient information to propose larger moves while maintaining reasonable acceptance probabilities. Its efficiency, however, depends on the step size $\epsilon$, the number of leapfrog steps $L$, and especially the mass matrix $\mathbf M$, or equivalently the inverse mass matrix $\mathbf M^{-1}$ that appears in the position update. Poor choices of $\mathbf M$ can lead to slow exploration or require very small step sizes for stable integration. This motivates the covariance-based preconditioning strategy developed in Section~\ref{sec:hmc-gvi}, where the target covariance estimate $\mathbf S$ is used as the inverse mass matrix $\mathbf M^{-1}$.
\subsection{Gaussian variational inference}\label{sec2.2}
GVI approximates the target density $\pi$ within a Gaussian variational family. Specifically, the variational distribution $q_{\bm{\lambda}}(\bm{\theta})$ is taken to be a multivariate Gaussian with mean $\bm{\mu}$ and covariance matrix $\bm{\Sigma}$, where $\bm{\lambda}$ collects the mean and the parameters used to represent the covariance or precision matrix. The variational parameters are obtained by minimizing the KL divergence from $q_{\bm{\lambda}}$ to the target density,
\[
\begin{aligned}
    \bm{\lambda}^{\star}
    &=
    \arg\min_{\bm{\lambda}}
    \mathrm{KL}
    \left\{
    q_{\bm{\lambda}}(\bm{\theta})
    \,\|\, 
    \pi(\bm{\theta})
    \right\}  \\
    &=
    \arg\max_{\bm{\lambda}}
    \left[
    \mathbb E_{q_{\bm{\lambda}}}
    \{\ell(\bm{\theta})\}
    +
    \mathcal H(q_{\bm{\lambda}})
    \right],
\end{aligned}
\]
where $\mathcal H(q_{\bm{\lambda}})=-\mathbb E_{q_{\bm{\lambda}}}\{\log q_{\bm{\lambda}}(\bm{\theta})\}$ is the entropy of the variational distribution. The second equality follows because $\ell(\bm{\theta})=\log\tilde{\pi}(\bm{\theta})$ differs from $\log\pi(\bm{\theta})$ only by an additive normalizing constant independent of $\bm{\lambda}$. Different GVI variants mainly differ in how $\bm{\Sigma}$, or equivalently the precision matrix, is parameterized. We briefly review several commonly used parameterizations and their computational motivations.

GVI with Cholesky decomposition (CGVI) parameterizes the covariance matrix through its Cholesky factorization, $\bm{\Sigma} = \mathbf{L}\mathbf{L}^{\top}$, where $\mathbf{L}$ is a lower triangular matrix with positive diagonal entries \citep{arXiv_tran2021practical}. This parameterization gives a flexible full covariance Gaussian approximation and enables the reparameterization
$\bm{\theta} = \bm{\mu} + \mathbf{L}\bm{z}, \bm{z}\sim \mathcal N(\mathbf 0,\mathbf I)$, which leads to pathwise gradient estimators with reduced Monte Carlo variance \citep{C_ICLR_kingma2014autoencoding}. However, because the number of covariance parameters grows quadratically with the dimension of $\bm{\theta}$, CGVI can become computationally burdensome in high-dimensional problems. To improve scalability, GVI with factor-decomposed covariance (FGVI) replaces the unrestricted covariance matrix with a low-rank-plus-diagonal representation, $\bm{\Sigma} = \mathbf{B}\mathbf{B}^{\top}+
\operatorname{diag}(\mathbf c^2)$, where $\mathbf{B}$ is a factor loading matrix and $\mathbf c$ contains positive diagonal scale parameters \citep{J_JCGS_ong2018gaussian}. This structure retains part of the dependence structure of the target through the low-rank term while substantially reducing the number of variational parameters. Another scalable alternative is GVI with sparse precision matrices, proposed by \citet{J_SAC_tan2018gaussian}. Instead of directly parameterizing the covariance matrix, this approach assumes a Gaussian variational distribution with sparse precision matrix $\bm{\Omega} =\bm{\Sigma}^{-1} =\mathbf{L}_{\Omega}\mathbf{L}_{\Omega}^{\top}$.
Sparsity in the Cholesky factor $\mathbf{L}_{\Omega}$ induces sparsity in the precision matrix, allowing the approximation to encode conditional independence structure. Detailed algorithms for these GVI variants can be found in \citet{arXiv_tran2021practical,J_JCGS_ong2018gaussian,J_SAC_tan2018gaussian}.
\subsection{Variational Auto-Encoders and Extensions for Multi-modal Distributions}\label{sec2.3}
VAEs provide a flexible class of latent-variable generative models for approximating complex data distributions. For a generic observation $\bm{y}$ and latent variable $\bm{z}$, the standard VAE specifies
\[
    p_{\bm{\eta}}(\bm{y})
    =
    \int
    p_{\bm{\eta}}(\bm{y}\mid \bm{z})
    p(\bm{z})
    \,d\bm{z},
\]
where $p(\bm{z})$ is usually chosen as a standard Gaussian prior and $p_{\bm{\eta}}(\bm{y}\mid \bm{z})$ is a decoder distribution parameterized by a neural network. Since the latent posterior $p_{\bm{\eta}}(\bm{z}\mid \bm{y})$ is generally intractable, VAEs introduce an amortized variational approximation $q_{\bm{\phi}}(\bm{z}\mid \bm{y})$, called the encoder, and maximize the evidence lower bound
\[
    \mathcal L(\bm{\eta},\bm{\phi};\bm{y})
    =
    \mathbb E_{q_{\bm{\phi}}(\bm{z}\mid \bm{y})}
    \left[
    \log p_{\bm{\eta}}(\bm{y}\mid \bm{z})
    \right]
    -
    \mathrm{KL}
    \left\{
    q_{\bm{\phi}}(\bm{z}\mid \bm{y})
    \,\|\, 
    p(\bm{z})
    \right\}.
\]
The reparameterization trick yields a pathwise gradient estimator for this objective, allowing the encoder and decoder parameters to be optimized jointly \citep{C_ICLR_kingma2014autoencoding}.

The basic VAE is computationally attractive, but its common use of a unimodal Gaussian prior and a factorized Gaussian variational posterior can be restrictive for distributions with separated modes. One line of work addresses this limitation by increasing the flexibility of the variational posterior over latent variables. For example, normalizing flows transform a simple base density through a sequence of invertible maps,
\begin{equation}
    \bm{z}_T
    =
    f_T\circ\cdots\circ f_1(\bm{z}_0),
    \qquad
    \log q_T(\bm{z}_T\mid\bm{y})
    =
    \log q_0(\bm{z}_0\mid\bm{y})
    -
    \sum_{t=1}^T
    \log
    \left|
    \det
    \frac{\partial f_t(\bm{z}_{t-1})}{\partial \bm{z}_{t-1}}
    \right|.
    \label{flow_vae}
\end{equation}
This construction gives a richer variational posterior while keeping density evaluation tractable through the change-of-variables formula \citep{C_ICML_rezende2015variational}. However, improving the posterior approximation alone does not necessarily solve the generative mismatch caused by a simple prior.

A complementary direction is to introduce mixture structure into the latent generative model. Let $c\in\{1,\ldots,K\}$ be a discrete latent component indicator. Together with the continuous latent variable $\bm{z}$, this gives the marginal model
\[
    p_{\bm{\eta},\bm{\rho}}(\bm{y})
    =
    \sum_{c=1}^K
    p(c)
    \int
    p_{\bm{\eta}}(\bm{y}\mid \bm{z},c)
    p_{\bm{\rho}}(\bm{z}\mid c)
    \,d\bm{z}.
\]
The corresponding inference model is often factorized as
\[
    q_{\bm{\phi}}(\bm{z},c\mid \bm{y})
    =
    q_{\bm{\phi}}(c\mid \bm{y})
    q_{\bm{\phi}}(\bm{z}\mid \bm{y},c).
\]
Under this factorization, the ELBO becomes
\[
\begin{aligned}
    \mathcal L(\bm{\eta},\bm{\rho},\bm{\phi};\bm{y})
    &=
    \mathbb E_{q_{\bm{\phi}}(\bm{z},c\mid \bm{y})}
    \left[
    \log p_{\bm{\eta}}(\bm{y}\mid \bm{z},c)
    \right]  \\
    &\quad
    -
    \mathrm{KL}
    \left\{
    q_{\bm{\phi}}(c\mid \bm{y})
    \,\|\, 
    p(c)
    \right\}  \\
    &\quad
    -
    \mathbb E_{q_{\bm{\phi}}(c\mid \bm{y})}
    \left[
    \mathrm{KL}
    \left\{
    q_{\bm{\phi}}(\bm{z}\mid \bm{y},c)
    \,\|\, 
    p_{\bm{\rho}}(\bm{z}\mid c)
    \right\}
    \right].
\end{aligned}
\]
The two regularization terms separate the assignment of observations to latent components from the regularization of the continuous latent representation within each component \citep{C_ECAI_lavda2020improving}. Related extensions include deep latent Gaussian mixture models, where mixture weights are treated as latent variables and handled through stick-breaking approximations \citep{C_NeurIPSW_nalisnick2016approximate}; mixtures of VAEs, where different VAE components are trained jointly with regularization encouraging the components to capture different parts of the data distribution \citep{C_IPTA_ye2020mixtures}; hierarchical decompositional mixtures, which combine VAE leaves with sum-product network structures to decompose high-dimensional density estimation into smaller local problems \citep{C_ICML_tan2019hierarchical}; and heavy-tailed decoder variants, such as Student-$t$ VAEs for robust density estimation, which replace the Gaussian decoder with a Student-$t$ distribution \citep{C_IJCAI_takahashi2018student}. In Section~\ref{sec:generative-proposal-family}, we use these ideas to construct VAE-based proposal families for the proposed Metropolis--Hastings sampler.

\section{Methodology}
\subsection{Linear Transformation in HMC via GVI}
\label{sec:hmc-gvi}

HMC is invariant under nonsingular linear transformations \citep{BS_neal2011mcmc}. This property provides useful guidance for choosing the momentum distribution when the target distribution has strong scale differences or correlations. Let $\tilde{\pi}(\bm{\theta})$ denote the unnormalized target density on $\mathbb R^d$, and let $\ell(\bm{\theta})=\log \tilde{\pi}(\bm{\theta})$. Suppose that the target is approximately Gaussian with covariance matrix $\bm{\Sigma}$. If $ \bm{\Sigma}=\mathbf L\mathbf L^{\top}$, then the whitening transformation $\bm{\eta}= \mathbf L^{-1}(\bm{\theta}-\bm{\mu})$ maps the approximately correlated distribution in the original $\bm{\theta}$-space to an approximately isotropic distribution in the transformed $\bm{\eta}$-space.

Running isotropic HMC in the whitened coordinates is equivalent to running HMC in the original coordinates with a non-isotropic momentum distribution. To see this, write $ \bm{\theta}=\bm{\mu}+\mathbf L\bm{\eta}$. The corresponding canonical transformation of the momentum variables is $\bm{\psi} = \mathbf L^{-\top}\bm{\rho}$, where $\bm{\rho}$ is the momentum variable in the whitened coordinate system. If $\bm{\rho}\sim \mathcal N(\mathbf 0,\mathbf I)$, then
$
    \bm{\psi}
    \sim
    \mathcal N
    \left(
    \mathbf 0,
    \mathbf L^{-\top}\mathbf L^{-1}
    \right)
    =
    \mathcal N
    \left(
    \mathbf 0,
    \bm{\Sigma}^{-1}
    \right).
$
The kinetic energy in the original coordinates is therefore
$
    K(\bm{\psi})
    =
    \frac12
    \bm{\psi}^{\top}
    \bm{\Sigma}
    \bm{\psi}.
$
Equivalently, the Hamiltonian can be written as
$
    H(\bm{\theta},\bm{\psi})
    =
    -\ell(\bm{\theta})
    +
    \frac12
    \bm{\psi}^{\top}
    \bm{\Sigma}
    \bm{\psi}.
$
The corresponding Hamiltonian equations are
$
    \frac{d\bm{\theta}}{dt}
    =
    \nabla_{\bm{\psi}}K(\bm{\psi})
    =
    \bm{\Sigma}\bm{\psi},
    \frac{d\bm{\psi}}{dt}
    =
    \nabla_{\bm{\theta}}\ell(\bm{\theta}).
$
Thus, when the momentum covariance is chosen as $\bm{\Sigma}^{-1}$, the leapfrog position update in the original parameterization is
$
    \bm{\theta}
    \leftarrow
    \bm{\theta}
    +
    \epsilon
    \bm{\Sigma}\bm{\psi}.
$
Therefore, the effect of whitening can be implemented without explicitly transforming the Markov chain to the $\bm{\eta}$-space: in the original parameterization, one may use the position covariance matrix as the inverse mass matrix, or preconditioning matrix, in the HMC position update.

In practice, the true target covariance is unknown. We therefore use GVI to obtain a Gaussian approximation
$
    q_{\mathrm{GVI}}(\bm{\theta})
    =
    \mathcal N
    \left(
    \bm{\mu}_{\mathrm{GVI}},
    \bm{\Sigma}_{\mathrm{GVI}}
    \right)
$
to the target distribution. The GVI approximation may be obtained from the Cholesky, factor, or sparse-precision parameterizations reviewed in Section~\ref{sec2.2}. In all cases, we write the implied covariance estimate as $\bm{\Sigma}_{\mathrm{GVI}}$. For sparse-precision GVI, the implied covariance can be recovered from the fitted precision representation and is used as $\mathbf S$ in our implementation; equivalently, implementations may use the precision factor directly when drawing the momentum or evaluating $\mathbf S^{-1}$-related quantities.

The GVI mean is used as the initial value of the Markov chain, and the GVI covariance is used as a fixed estimate of the position covariance,
$
    \mathbf S
    =
    \bm{\Sigma}_{\mathrm{GVI}}.
$
The matrix $\mathbf S$ is used as the inverse mass matrix, $\mathbf S=\mathbf M^{-1}$, so the corresponding momentum covariance is $\mathbf M=\mathbf S^{-1}$. We then sample the momentum from
$
    \bm{\psi}
    \sim
    \mathcal N(\mathbf 0,\mathbf S^{-1})
$
and use the kinetic energy
$
    K(\bm{\psi})
    =
    \frac12
    \bm{\psi}^{\top}\mathbf S\bm{\psi}.
$
For any positive definite matrix $\mathbf S$ fixed during the main sampling stage, the Metropolis--Hastings correction preserves the target distribution. The accuracy of $\mathbf S$ affects the efficiency of the sampler, but not the validity of the resulting Markov chain.

The resulting GVI-preconditioned HMC algorithm is given in Algorithm~\ref{alg:hmc-gvi}.

For comparison, we also consider a burn-in-based covariance estimate. In that version, a preliminary HMC run is first performed with a simple choice such as $\mathbf S=\mathbf I_d$. The empirical covariance matrix of the preliminary samples is then used as
$
    \mathbf S_{\mathrm{burn}}
    =
    \widehat{\operatorname{Cov}}
    \left(
    \bm{\theta}^{(1)},\ldots,\bm{\theta}^{(B_0)}
    \right).
$
The main HMC run is subsequently performed using $\mathbf S=\mathbf S_{\mathrm{burn}}$ in Algorithm~\ref{alg:hmc-gvi}. In both the GVI-based and burn-in-based versions, the matrix $\mathbf S$ is fixed during the main sampling stage. The step size $\epsilon$ and the number of leapfrog steps $L$ are selected by short preliminary runs, with the aim of obtaining stable acceptance probabilities and reducing autocorrelation in the resulting Markov chain.

\subsection{Generative-Proposal Metropolis-Hastings for Multi-modal Targets}\label{sec3.2}

Sampling from a multi-modal target density $\pi(\bm{\theta})$ is difficult for three related reasons. First, the sampler must identify high-probability regions where the modes are located. Second, it must move between these regions by crossing low-probability barriers. Third, once a mode has been reached, it must sample efficiently within that local region while accounting for differences in scale, shape, and local geometry. We address these three tasks using a generative-proposal MH sampler. The resulting method combines a mode-finding stage, a VAE-based global proposal, a kernel-density MH correction, and a local random-walk kernel.

\subsubsection{Identifying high-probability regions}
Before training a generative proposal, we need an initial set of samples that covers the important regions of the target. We use Crowding-based Differential Evolution (CrowdingDE)~\citep{C_CEC_thomsen2004multimodal}, a population-based multi-modal optimization method, to identify approximate mode locations. The complete pseudo-code is provided in Algorithm~\ref{alg:differential_evolution}. Conceptually, CrowdingDE searches for local maximizers of the unnormalized target density $\tilde{\pi}(\bm{\theta})$. In the implementation, this is equivalently done by minimizing $-\tilde{\pi}(\bm{\theta})$. This approach is useful because the population can maintain multiple promising regions simultaneously, whereas repeated runs of a standard local optimizer may still return duplicated modes. For further details, please refer to~\cite{J_JGO_storn1997differential} and the recent survey~\citep{J_AIR_chauhan2025advancements}.


Let $\widehat{\bm{m}}_1,\ldots,\widehat{\bm{m}}_J$ denote the mode locations returned by the CrowdingDE step above. These estimated modes are then used to generate an initial training set
\begin{equation*}
    \mathcal{D}_0
    =
    \bigcup_{j=1}^{J}
    \left\{
    \bm{\theta}_{j\ell}:
    \begin{array}{l}
    \bm{\theta}_{j\ell}\sim \mathcal{N}(\widehat{\bm{m}}_j,\mathbf{I}_d),
    \ell=1,\ldots,n_0
    \end{array}
    \right\},
\end{equation*}
where $n_0$ is the number of initial samples generated around each estimated mode. This initialization gives the generative model information about all relevant high-probability regions before the MCMC chain begins.
\subsubsection{Learning a generative proposal family}
\label{sec:generative-proposal-family}

Let $\tau$ be the number of MH updates between two proposal refreshes and let $W$ be the training-window size. The sampler starts from a random point in $\mathcal{D}_0$ and uses $\mathcal{D}_0$ itself as the initial proposal pool. The implementation keeps a training history $\mathcal H_r$, initialized by $\mathcal D_0$ and then augmented by the subsequent MH states. After each block of $\tau$ MH updates, the VAE-type proposal is retrained using the most recent elements of this history. Specifically, at refresh round $r$, the training set is
$
    \mathcal D_r
    =
    \operatorname{last}_W(\mathcal H_r),
    |\mathcal D_r|\leq W,
$
where $\operatorname{last}_W(\mathcal H_r)$ denotes the most recent $W$ stored states, or all stored states if fewer than $W$ are available. The initial points in $\mathcal D_0$ are used for proposal learning and are discarded together with burn-in when computing final Monte Carlo summaries. In this step, the parameter vector $\bm{\theta}$ is treated as the observation variable in a VAE-type generative model. The fitted model is then used to generate global proposal candidates directly in the original parameter space.

We consider four VAE-based proposal families. The first is the standard VAE, which uses
$
    p(\bm{z})=\mathcal N(\bm{0},\mathbf{I}),
    q_{\bm{\phi}}(\bm{z}\mid\bm{\theta})
    =
    \mathcal N
    \left(
    \bm{\mu}_{\bm{\phi}}(\bm{\theta}),
    \operatorname{diag}\{
    \bm{\sigma}^2_{\bm{\phi}}(\bm{\theta})
    \}
    \right),
$
together with a decoder $p_{\bm{\eta}}(\bm{\theta}\mid\bm{z})$. After training, proposal candidates are generated by ancestral sampling: draw $\bm{z}\sim\mathcal N(\bm{0},\mathbf{I})$ and decode it to obtain $\bm{\theta}^{\star}$. This serves as the baseline generative proposal. The second proposal family is a flow-augmented VAE. It replaces the diagonal Gaussian encoder with a normalizing-flow posterior of the form in \eqref{flow_vae}. In our implementation, the flow transformation can be chosen from RealNVP affine coupling layers \citep{C_ICLR_dinh2017density}, masked autoregressive flows \citep{C_NeurIPS_papamakarios2017masked}, planar flows \citep{C_ICML_rezende2015variational}, or neural spline flows \citep{C_NeurIPS_durkan2019neural}. These transformations are used on the encoder side during training, thereby increasing the flexibility of the variational posterior and improving the learned decoder when the training samples exhibit nonlinear dependence. At the proposal-generation stage, candidates are still produced by sampling from the latent prior and decoding to the parameter space. The third proposal family is a component-conditioned GMM-VAE, designed for targets with separated modes. It introduces a discrete component indicator $c\in\{1,\ldots,K\}$ and specifies the latent generative model as
\begin{equation}
    c\sim p_{\bm{\rho}}(c),
    \qquad
    \bm{z}\mid c
    \sim
    p_{\bm{\rho}}(\bm{z}\mid c),
    \qquad
    \bm{\theta}\mid \bm{z},c
    \sim
    p_{\bm{\eta}}(\bm{\theta}\mid \bm{z},c).
    \label{eq:gmm_vae_prior}
\end{equation}
The inference model factorizes as
$
    q_{\bm{\phi}}(\bm{z},c\mid \bm{\theta})
    =
    q_{\bm{\phi}}(c\mid \bm{\theta})
    q_{\bm{\phi}}(\bm{z}\mid \bm{\theta},c).
$
The decoder is conditioned on both the continuous latent variable $\bm{z}$ and the component label $c$. This structure allows different components to specialize to different regions of the target distribution and is therefore well suited to multi-modal sampling. In the implementation, the component centers may be initialized from the estimated mode locations, and component-balanced generation can be used to prevent small modes from being underrepresented in the proposal pool.

Finally, for heavy-tailed targets, we replace the Gaussian decoder likelihood with a Student $t$ decoder likelihood. Specifically, the reconstruction model is taken to be
\begin{equation*}
    p_{\bm{\eta}}(\bm{\theta}\mid \bm{z},c)
    =
    t_{\nu}
    \left(
    \bm{\theta}
    \mid
    \bm{m}_{\bm{\eta}}(\bm{z},c),
    s^2\mathbf{I}_d
    \right),
\end{equation*}
with fixed degrees of freedom $\nu$ and scale $s$; for standard or flow VAEs without a component label, the conditioning on $c$ is omitted. This modification is used when the target has heavier tails than a Gaussian mixture, since it reduces the mismatch between the decoder likelihood and the empirical tail behavior of the training samples.

Across all variants, the learned VAE-type model is used only as a mechanism for generating global candidates. The quality of the generative proposal affects mixing efficiency and mode exploration, while the accept-reject correction is handled by the KDE-based MH step described below.

\subsubsection{Moving between modes with KDE-corrected global proposals}

Let $\mathcal P_r$ denote the proposal pool used during round $r$. The initial pool is $\mathcal P_0=\mathcal D_0$. For $r\geq 1$, after training a generative model $G_r$ on $\mathcal{D}_r$, we draw
$
    \mathcal P_r=\{\bm{\zeta}_{r,1},\ldots,\bm{\zeta}_{r,M}\},
    \bm{\zeta}_{r,m}\sim G_r.
$
Because the marginal proposal density induced by sampling latent variables and decoding them is generally not available in closed form, we construct an explicit proposal density by applying kernel density estimation (KDE) to the current proposal pool:
\begin{equation}
    q_r(\bm{\theta})=
    \frac{1}{M}\sum_{m=1}^{M}
    \varphi_h(\bm{\theta}-\bm{\zeta}_{r,m}),
    \qquad
    \varphi_h(\bm{v})=
    \mathcal N(\bm{v}\mid \bm{0},h^2\mathbf{I}_d),
    \label{eq:kde_proposal}
\end{equation}
where $h$ is the KDE bandwidth. The global proposal is independent of the current state: to propose $\bm{\theta}'$, we select $I\sim U\{1,\ldots,M\}$ and draw
$
    \bm{\theta}'=\bm{\zeta}_{r,I}+\bm{\epsilon}_g,
    \bm{\epsilon}_g\sim\mathcal N(\bm{0},h^2\mathbf{I}_d).
$
The proposal density of this move is exactly $q_r$, so the MH acceptance probability is
\begin{equation}
    \alpha_g(\bm{\theta},\bm{\theta}')
    =
    \min\left\{
    1,
    \frac{\pi(\bm{\theta}')q_r(\bm{\theta})}
    {\pi(\bm{\theta})q_r(\bm{\theta}')}
    \right\}.
    \label{eq:vae_mh_acceptance}
\end{equation}
The normalizing constant of $\pi$ is not needed because it cancels in the ratio. This correction is important: the VAE is used to construct a useful global proposal, while the MH step corrects for the discrepancy between the learned proposal and the true target density. Conditional on the past history and hence on a fixed proposal density $q_r$ within a refresh round, the global MH kernel satisfies detailed balance with respect to $\pi$.

\subsubsection{Sampling efficiently within modes}

A global proposal is useful for jumping between modes, but it is not always the most efficient mechanism for local exploration. We therefore combine it with a Gaussian random-walk proposal. At iteration $t$, with probability $\rho$ we use the KDE-corrected global proposal in equation~\eqref{eq:vae_mh_acceptance}; with probability $1-\rho$, we propose
$
    \bm{\theta}'=\bm{\theta}^{(t)}+\bm{\epsilon},
    \bm{\epsilon}\sim\mathcal{N}(\bm{0},\sigma_{\mathrm{rw}}^2\mathbf{I}_d),
$
and accept it with the usual random-walk MH probability
$
    \alpha_{\mathrm{rw}}(\bm{\theta},\bm{\theta}')
    =
    \min\left\{1,\frac{\pi(\bm{\theta}')}{\pi(\bm{\theta})}\right\}.
$
The local kernel improves exploration within individual modes, while the global kernel prevents the chain from being trapped in one mode. For a fixed proposal pool $\mathcal P_r$, both the KDE-corrected global kernel and the symmetric random-walk kernel are reversible with respect to $\pi$, and their mixture is therefore also reversible with respect to $\pi$. Since the proposal pool is updated across rounds using past states, the full procedure is an adaptive MCMC scheme rather than a single time-homogeneous MH kernel. Thus, the detailed-balance statement applies within each fixed-pool round, during which the proposal density is held fixed for the $\tau$ MH updates. The complete procedure is summarized in Algorithm~\ref{VAE-MH}.

\section{Numerical Studies}\label{sec5}
\subsection{Linear Transformations in HMC via GVI}
We evaluate HMC with GVI-derived linear transformations. For the logistic regression and Gaussian examples, we compare five MCMC algorithms: random-walk MH with scale matrix estimated from burn-in samples (RMH), AM, preconditioned MALA, HMC with a linear transformation matrix estimated from burn-in samples (HMC), and HMC with a linear transformation matrix estimated via GVI (HMC-GVI). For each algorithm, we employ standard hyper-parameter tuning procedures to ensure fair comparison. Each algorithm runs $10^6$ iterations with $10^4$ burn-in iterations. Each result is an average over 3 replicated runs. Throughout this subsection, $E$ denotes the normalized mean effective sample size, computed as the mean ESS divided by the number of retained samples.
\subsubsection{Posterior sampling in logistic regression models}
Let $\mathbf{y}=(y_1,\ldots,y_n)^{\top}$ be a vector of observed binary responses and $\mathbf{X}$ the corresponding design matrix with rows $\mathbf{x}_{i}=(1,x_{i2},\ldots,x_{id})^{\top}$ for $i=1,2,\ldots,n$. All non-intercept predictors are standardized to improve mixing. The generalized linear model is
$
    (y_i\mid p_i)\overset{ind}{\sim} \text{Bernoulli}(p_i), p_i = g(\bm{\beta}^{\top}\mathbf{x}_{i}),
$
where $g(\cdot)$ is the inverse logit transform, i.e., $g(\eta) = e^{\eta}/(1 + e^{\eta})$. The log-likelihood function is:
$
    \ell_{\mathrm{lik}}(\bm{\beta}) = \sum_{i=1}^{n}\left[y_{i} \bm{\beta}^{\top} \mathbf{x}_{i}-\log \left(1+e^{\bm{\beta}^{\top} \mathbf{x}_{i}}\right)\right].
$
We specify a vague prior $\bm{\beta} \sim \mathcal{N}_d(\mathbf{0},100\mathbf{I}_d)$, yielding the posterior target:
$
    \pi(\bm{\beta}) \equiv \pi(\bm{\beta} \mid \mathbf{y}) \propto \exp\{\ell_{\mathrm{lik}}(\bm{\beta})\}p(\bm{\beta})
$. Two datasets are used: the Pima Indian dataset ($d=8$, $n=532$) and the German credit dataset ($d=25$, $n=1000$).

For these moderate-dimensional problems, we use CGVI to derive Gaussian approximations for the HMC linear transformation matrices. The CGVI running times (2.5434s and 3.6831s) are substantially shorter than the time used by the preliminary HMC run for estimating the transformation matrix (82.5673s and 125.9163s).

\begin{table}[htbp]
    \centering
    \caption{Comparison of five MCMC algorithms on Pima Indian dataset ($d=8$, $n=532$).}
    \label{PIMA}
    \setlength{\tabcolsep}{0.9mm}{
        \begin{tabular}{lcccc}
            \toprule
            Method  & Time(s)  & $E$             & $P_{\text{jump}}$ & $\rho_1$ \\ \midrule
            RMH     & 17.3823  & 0.0371          & 0.2512            & 0.9253   \\
            AM      & 37.9495  & 0.0377          & 0.2501            & 0.9252   \\
            MALA    & 51.4006  & 0.2599          & 0.4987            & 0.5510   \\
            HMC     & 580.4651 & 0.9716          & 0.9773            & -0.0066  \\
            HMC-GVI & 345.2453 & \textbf{0.9972} & 0.9929            & -0.0223  \\
            \bottomrule
        \end{tabular}}
\end{table}

\begin{table}[htbp]
    \centering
    \caption{Comparison of five MCMC algorithms on German credit dataset ($d=25$, $n=1000$).}
    \label{GERMAN}
    \setlength{\tabcolsep}{0.9mm}{
        \begin{tabular}{lcccc}
            \toprule
            Method  & Time(s)  & $E$             & $P_{\text{jump}}$ & $\rho_1$ \\ \midrule
            RMH     & 22.9772  & 0.0110          & 0.2405            & 0.9744   \\
            AM      & 47.0808  & 0.0108          & 0.2293            & 0.9743   \\
            MALA    & 69.5715  & 0.1469          & 0.5718            & 0.7283   \\
            HMC     & 938.6637 & 0.9105          & 0.9752            & 0.0037   \\
            HMC-GVI & 404.6909 & \textbf{0.9911} & 0.9118            & -0.0274  \\
            \bottomrule
        \end{tabular}}
\end{table}

Tables \ref{PIMA} and \ref{GERMAN} present performance metrics averaged over all parameters. Results show that the GVI-derived transformation improves efficiency with less preprocessing overhead: $E = 0.9972$ and $E = 0.9911$ for HMC-GVI versus $E = 0.9716$ and $E = 0.9105$ for standard HMC.

\begin{figure}[H]
    \begin{minipage}[b]{0.5\linewidth}
        \centering
        \includegraphics[width=0.95\linewidth]{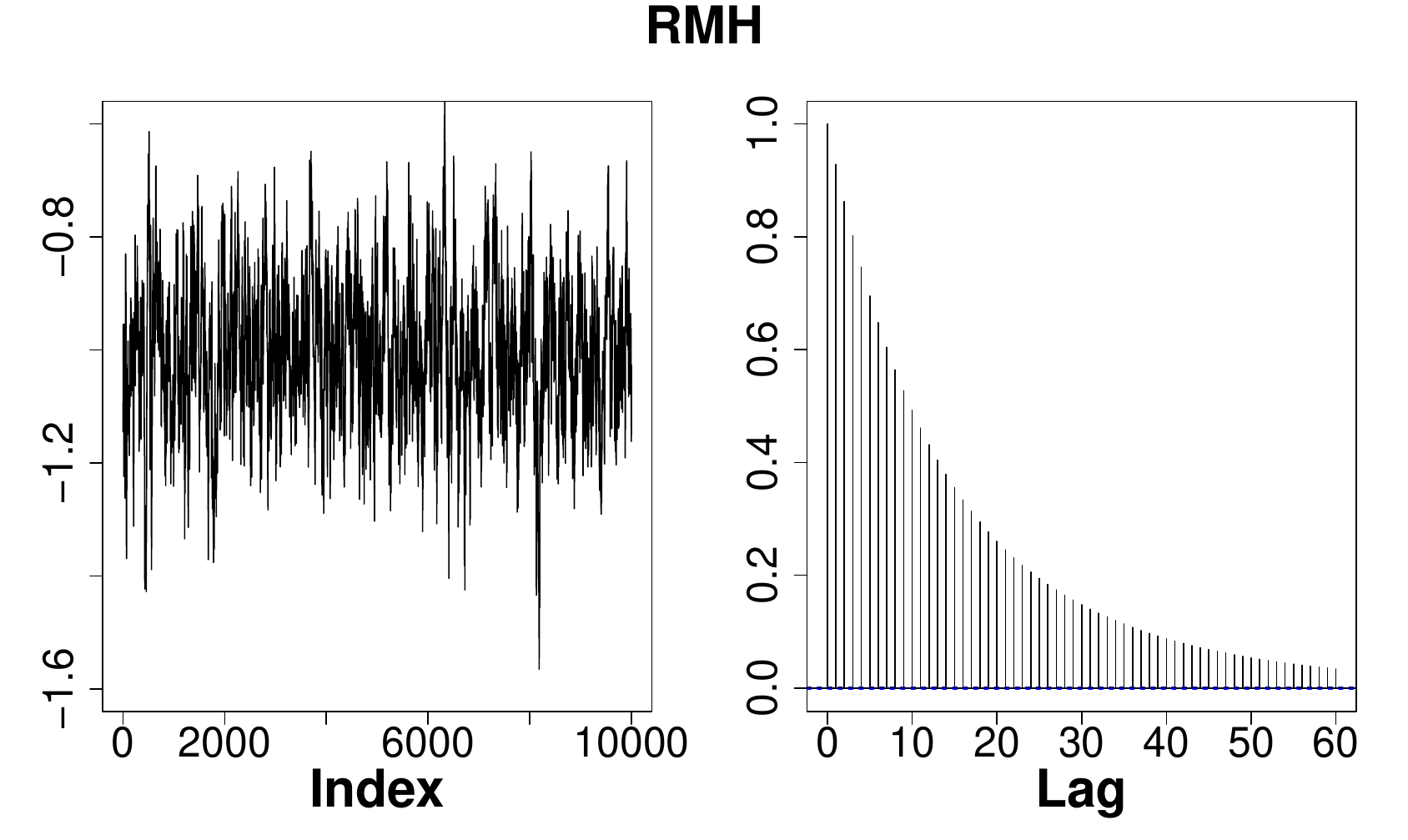}
    \end{minipage}%
    \begin{minipage}[b]{0.5\linewidth}
        \centering
        \includegraphics[width=0.95\linewidth]{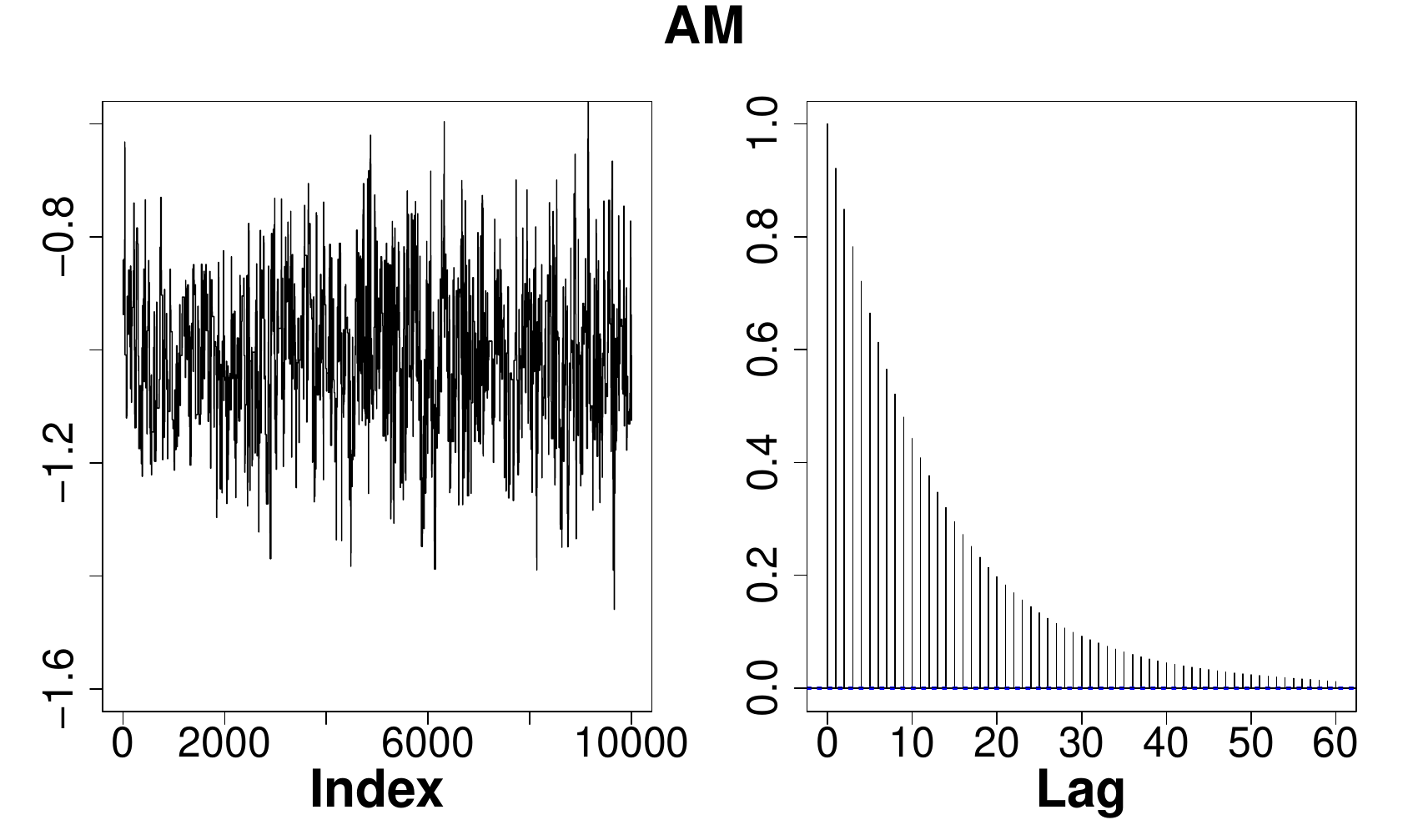}
    \end{minipage}\\[6pt]
    \begin{minipage}[b]{0.5\linewidth}
        \centering
        \includegraphics[width=0.95\linewidth]{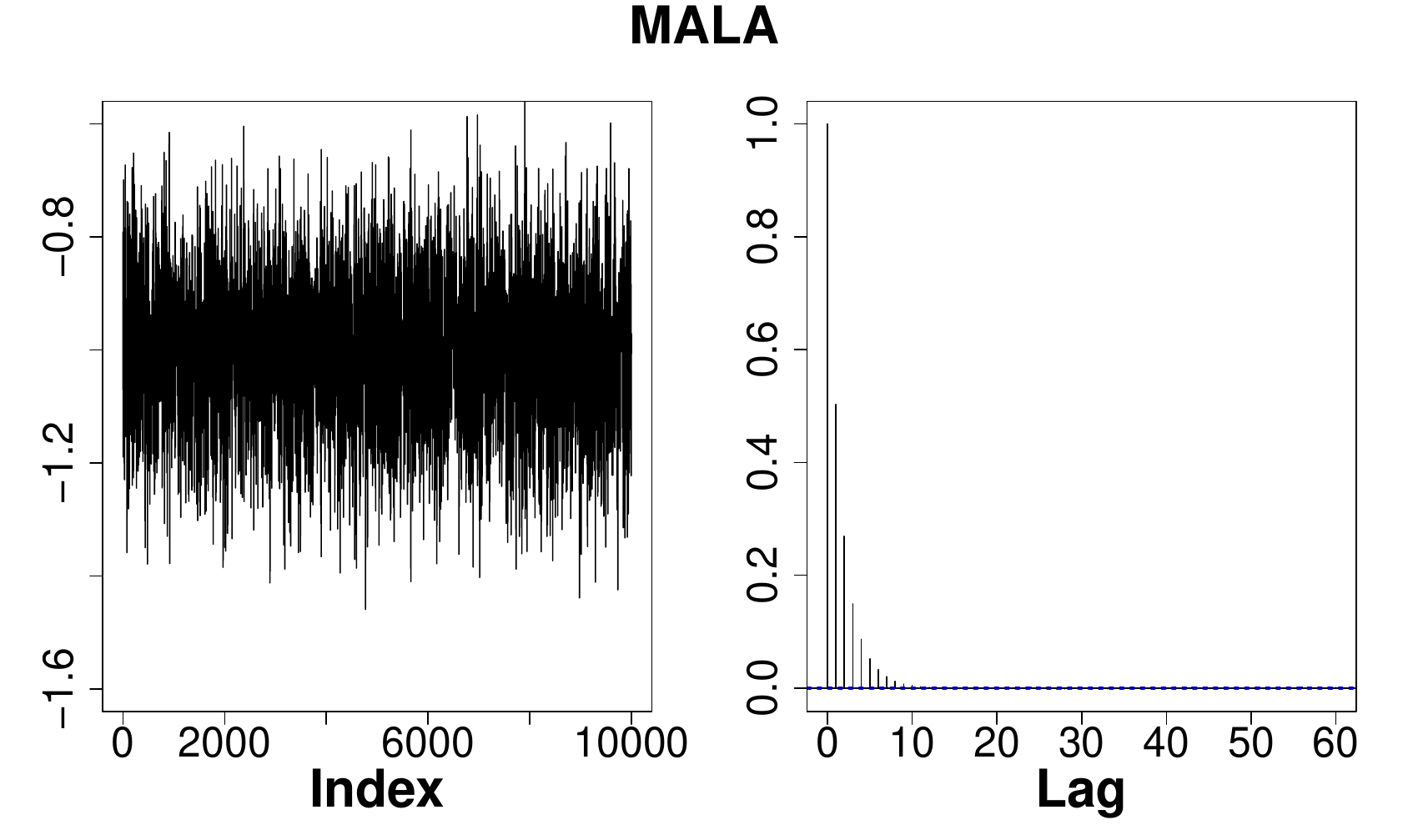}
    \end{minipage}%
    \begin{minipage}[b]{0.5\linewidth}
        \centering
        \includegraphics[width=0.95\linewidth]{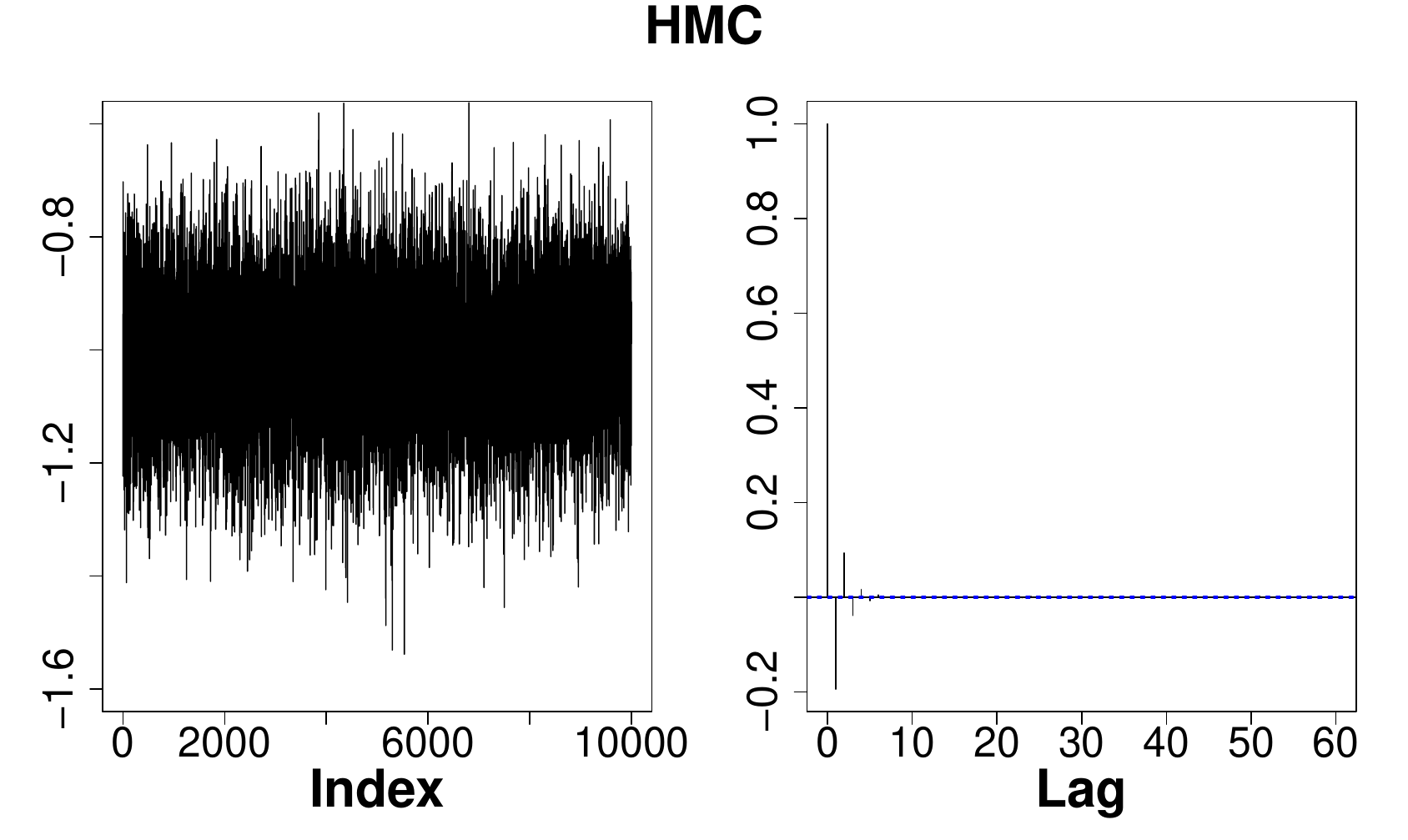}
    \end{minipage}\\[6pt]
    \centering
    \includegraphics[width=0.5\linewidth]{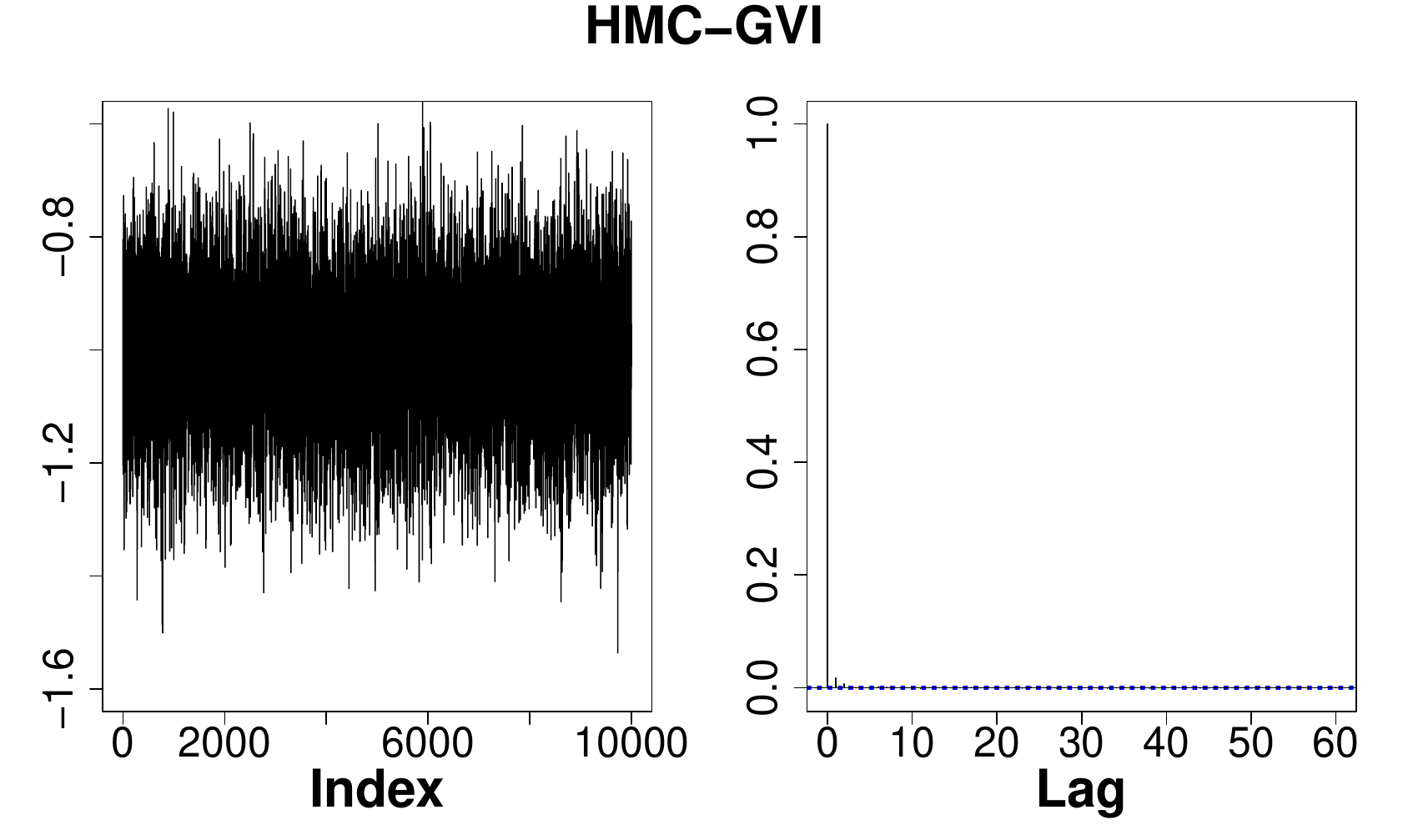}
    \caption{Trace and ACF plots of $\bm{\beta}_1$ for five MCMC algorithms on Pima Indian dataset.}
    \label{fig:composite}
\end{figure}

Figure \ref{fig:composite} displays trace and ACF plots for $\bm{\beta}_1$ in the Pima Indian dataset. MALA, HMC, and HMC-GVI demonstrate superior mixing compared to RMH and AM, with faster autocorrelation decay. HMC-GVI shows the best performance, achieving near-optimal efficiency and minimal autocorrelation.

\subsubsection{Sampling from a 100-dimensional Gaussian distribution}

We next consider a 100-dimensional Gaussian target distribution,
\[
    \pi_{100}(\bm{\theta}) \propto
    \exp\left\{
    -\frac{1}{2}\bm{\theta}^{\top}
    \bm{\Sigma}^{-1}
    \bm{\theta}
    \right\},
    \qquad
    \bm{\theta}\in\mathbb{R}^{100},
\]
where the mean vector is zero and the covariance matrix is denoted by $\bm{\Sigma}$. The diagonal entries of $\bm{\Sigma}$ are independently generated from a Gamma distribution with shape parameter $2$ and scale parameter $3$, while the off-diagonal entries are set to $0.8$. Only positive definite realizations of $\bm{\Sigma}$ are retained. The corresponding log-density, up to an additive constant, is given by
$
    \ell(\bm{\theta})
    =
    -\frac{1}{2}
    \bm{\theta}^{\top}
    \bm{\Sigma}^{-1}
    \bm{\theta},
$
with gradient
$
    \nabla_{\bm{\theta}}\ell(\bm{\theta})
    =
    -\bm{\Sigma}^{-1}\bm{\theta}.
$

For the HMC algorithm with linear transformation via GVI, we employ FGVI with rank $f=5$ to compute the covariance matrix. The average running time of FGVI is 5.1956s, while the preliminary HMC covariance-estimation stage takes 52.6496s. The running time, efficiency ($E$), average acceptance rate ($P_{\text{jump}}$) and lag-1 autocorrelation ($\rho_1$) are shown for each algorithm in Table~\ref{tab:comparison_100d_gaussian}.
\begin{table}[htbp]
    \centering
    \caption{Comparison of five MCMC algorithms for sampling a 100-dimensional Gaussian distribution.
        $E$: normalized mean effective sample size;
        $P_{\text{jump}}$: Average jump acceptance rate;
        $\rho_1$: Average lag-1 autocorrelation over coordinates.}
    \label{tab:comparison_100d_gaussian}
    \begin{tabular}{lcccc}
        \toprule
        Method  & {Time (s)} & {$E$}             & {$P_{\text{jump}}$} & {$\rho_1$} \\
        \midrule
        RMH     & 8.8393     & $2.52\times 10^{-5}$ & 0.5593         & 0.9972     \\
        AM      & 117.7322   & 0.0025            & 0.3771              & 0.9938     \\
        MALA    & 24.7687    & 0.0049            & 0.5675              & 0.9620     \\
        HMC     & 297.8293   & 0.9372            & 0.9744              & -0.0022    \\
        HMC-GVI & 213.0584   & \textbf{0.9884}   & 0.9744              & -0.0067    \\
        \bottomrule
    \end{tabular}
\end{table}

The results in Table~\ref{tab:comparison_100d_gaussian} show that for this high-dimensional target distribution, the running time of FGVI is significantly shorter than that of the preliminary HMC run used for covariance estimation. In the matched HMC comparison, replacing the preliminary-sample covariance estimate by the FGVI-derived matrix increases the normalized efficiency from $E = 0.9372$ to $E = 0.9884$, with comparable acceptance rates and lag-1 autocorrelation. Thus, in this example, the GVI-based transformation provides an efficient preconditioner while avoiding the more expensive preliminary MCMC covariance-estimation stage.

\subsubsection{Posterior sampling in a 509-dimensional GLMM}

We finally consider posterior sampling for a high-dimensional generalized linear mixed model (GLMM). Let $y_{ij}$ denote the binary response for subject $i=1,\ldots,n$ at occasion $j=1,\ldots,n_i$. Conditional on the subject-specific random effect $b_i$, we assume
$
    y_{ij}\mid b_i,\bm{\beta} \sim \mathrm{Bernoulli}(\mu_{ij}),
    \mathrm{logit}(\mu_{ij})
    =
    X_{ij}^{\top}\bm{\beta}
    +
    Z_{ij}^{\top}b_i ,
$
where $\bm{\beta}$ denotes the fixed-effect coefficients and $b_i$ denotes the random effect for subject $i$. The random effects are assumed to satisfy
$
    b_i\mid \bm{\xi} \sim N(\mathbf 0,\mathbf G),
    \mathbf G=\mathbf L\mathbf L^{\top},
$
where $\mathbf L$ is the Cholesky factor of $\mathbf G$. To ensure an unconstrained parametrization, we work with the transformed Cholesky factor $\mathbf L^\ast$, defined by
$
    L_{kk}^\ast=\log(L_{kk}), L_{kl}^\ast=L_{kl}\quad (k\ne l),
$
and set
$
    \bm{\xi}=\operatorname{vech}(\mathbf L^\ast).
$
The priors are specified as
$
    \bm{\beta}\sim N(\mathbf 0,\sigma_\beta^2\mathbf I),
    \bm{\xi}\sim N(\mathbf 0,\sigma_\xi^2\mathbf I).
$

Let
$
    \bm{\theta}
    =
    (\bm{\beta}^{\top}, b_1^{\top},\ldots,b_n^{\top},
    \bm{\xi}^{\top})^{\top}
$
denote all unknown parameters. Up to a normalizing constant, the posterior target is
\[
    \pi(\bm{\theta}) \equiv \pi(\bm{\theta}\mid \mathbf y)
    \propto
    p(\bm{\beta})p(\bm{\xi})
    \prod_{i=1}^{n}
    \left\{
    p(b_i\mid \bm{\xi})
    \prod_{j=1}^{n_i}
    p(y_{ij}\mid \bm{\beta},b_i)
    \right\}.
\]
Equivalently, the log-posterior density, up to an additive constant, is
\begin{equation}
    \begin{aligned}
        \log \pi(\bm{\theta})
        =
         & \sum_{i=1}^{n}\sum_{j=1}^{n_i}
        \left[
            y_{ij}\eta_{ij}
            -
            h_1(\eta_{ij})
            \right]
        -
        n\log|\mathbf L|
        -\frac{1}{2}\sum_{i=1}^{n}
        b_i^{\top}\mathbf L^{-\top}\mathbf L^{-1}b_i \\
         & -
        \frac{1}{2\sigma_\beta^2}\bm{\beta}^{\top}\bm{\beta}
        -
        \frac{1}{2\sigma_\xi^2}\bm{\xi}^{\top}\bm{\xi},
        \label{glmm-t}
    \end{aligned}
\end{equation}
where
$
    \eta_{ij}=X_{ij}^{\top}\bm{\beta}+Z_{ij}^{\top}b_i,
    h_1(x)=\log\{1+\exp(x)\}.
$

The gradient of the log-posterior is required for the implementation of gradient-based samplers. For $i=1,\ldots,n$,
\[
    \nabla_{b_i}\log \pi(\bm{\theta})
    =
    \sum_{j=1}^{n_i}
    \left\{
    y_{ij}-h_1'(\eta_{ij})
    \right\}Z_{ij}
    -
    \mathbf L^{-\top}\mathbf L^{-1}b_i,
\]
and
\[
    \nabla_{\bm{\beta}}\log \pi(\bm{\theta})
    =
    \sum_{i=1}^{n}\sum_{j=1}^{n_i}
    \left\{
    y_{ij}-h_1'(\eta_{ij})
    \right\}X_{ij}
    -
    \frac{1}{\sigma_\beta^2}\bm{\beta}.
\]
The gradient with respect to $\bm{\xi}$ is
\[
    \nabla_{\bm{\xi}}\log \pi(\bm{\theta})
    =
    -n\mathbf 1_{\mathrm{diag}(\mathbf L)}
    +
    \mathbf 1_{\bm{\xi}}\odot \operatorname{vech}(\mathbf A)
    -
    \frac{1}{\sigma_\xi^2}\bm{\xi},
\]
where
$
    \mathbf A
    =
    \sum_{i=1}^{n}
    \mathbf L^{-\top}\mathbf L^{-1}
    b_i b_i^{\top}
    \mathbf L^{-\top}.
$
Here, the $k$th element of $\mathbf 1_{\bm{\xi}}$ is $\exp(\xi_k)$ if $\xi_k$ corresponds to a diagonal element of $\mathbf L$, and is one otherwise. Similarly, the $k$th element of $\mathbf 1_{\mathrm{diag}(\mathbf L)}$ is one if $\xi_k$ corresponds to a diagonal element of $\mathbf L$, and is zero otherwise.

We use the polypharmacy dataset from the R package \texttt{aplore3}. Following \citet{J_SAC_tan2018gaussian}, we fit the logistic random-intercept model
$$
    \begin{aligned}
        \mathrm{logit}(\mu_{ij})
        =
         & \ \beta_0
        +\beta_1\mathrm{Gender}_i
        +\beta_2\mathrm{Race}_i
        +\beta_3\mathrm{Age}_{ij}
        +\beta_4\mathrm{MHV1}_{ij}    \\
         & +\beta_5\mathrm{MHV2}_{ij}
        +\beta_6\mathrm{MHV3}_{ij}
        +\beta_7\mathrm{INPTMHV}_{ij}
        +b_i ,
    \end{aligned}
$$
where $i=1,\ldots,500$ and $j=1,\ldots,7$. The response indicates whether the subject used drugs from three or more distinct groups. The covariates are gender, race, age, indicators for outpatient mental-health visits $\mathrm{MHV1}$, $\mathrm{MHV2}$, and $\mathrm{MHV3}$, and an indicator for inpatient mental-health visits $\mathrm{INPTMHV}$. Specifically, Gender is coded as one for male and zero for female, Race is coded as zero for white and one for other races, $\mathrm{MHV1}=1$ indicates one to five outpatient mental-health visits, $\mathrm{MHV2}=1$ indicates six to fourteen visits, $\mathrm{MHV3}=1$ indicates fifteen or more visits, and $\mathrm{INPTMHV}=1$ indicates at least one inpatient mental-health visit.

In this application, there are eight fixed-effect coefficients and 500 subject-specific random intercepts. Since the random-effect covariance reduces to a scalar variance component, $\bm{\xi}$ is one-dimensional. Thus, the full parameter vector has dimension $8+500+1=509$. We assign the priors
$
    \bm{\beta}\sim N(\mathbf 0,100\mathbf I_8),
    \xi\sim N(0,100),
    b_i\mid \xi \sim N(0,\exp(2\xi)).
$
The log-posterior target is then given by \eqref{glmm-t}.

To sample from the posterior distribution of the GLMM for the polypharmacy dataset, we compare the performance of four algorithms: preconditioned MALA with the preconditioning matrix estimated from preliminary MALA samples, preconditioned MALA with the matrix estimated via Gaussian variational inference (GVI), HMC with a linear transformation matrix estimated from preliminary HMC samples, and HMC with the matrix estimated via GVI. We exclude the random-walk MH and AM algorithms due to their low sampling efficiency. For each algorithm, we tune the hyper-parameters following the guidelines in the MCMC literature to achieve optimal sampling efficiency, and run a Markov chain of $10^6$ iterations with a burn-in period of $10^4$.

Due to the complex nature of the target distribution in this experiment, we use GVI with sparse precision matrices proposed by \citet{J_SAC_tan2018gaussian} to estimate the covariance structure used in both the MALA and HMC algorithms. For the polypharmacy dataset, the mean running time of sparse-precision GVI is approximately 5.6--6.2s across the GVI-based algorithms. Because this target is more complex than the preceding examples, when estimating the covariance matrix from preliminary MCMC samples, the preliminary covariance-estimation stages take around 41.6441s for MALA and 333.8184s for HMC on average.

The running time, efficiency ($E$), average acceptance rate ($P_{\text{jump}}$) and lag-1 autocorrelation ($\rho_1$) are shown for each algorithm in Table~\ref{tab:GLMM}. Each result is an average over 3 replicated runs.
\begin{table}[htbp]
    \centering
    \caption{Comparison of four MCMC algorithms on the polypharmacy dataset under the 509-dimensional GLMM model.}
    \label{tab:GLMM}

    \begingroup
    \setlength{\tabcolsep}{4pt}
    \renewcommand{\arraystretch}{1.1}

    \begin{tabular}{lcccc}
        \toprule
        Method   & Time (s)   & $E$             & $P_{\mathrm{jump}}$ & $\rho_1$ \\
        \midrule
        MALA     & 356.5509   & 0.0002          & 0.6802              & 0.9903   \\
        MALA-GVI & 319.1607   & 0.0267          & 0.6901              & 0.9424   \\
        HMC      & 7246.0966  & 0.8083          & 0.9772              & -0.0170  \\
        HMC-GVI  & 6112.1047  & \textbf{0.8762} & 0.9887              & -0.0106  \\
        \bottomrule
    \end{tabular}
    \endgroup
\end{table}
From Table~\ref{tab:GLMM}, we can see that for sampling from high-dimensional and complex target distributions, deriving the transformation matrix in the MALA and HMC algorithms using GVI is beneficial. First, the running time of GVI is significantly shorter than that of the preliminary MCMC runs.
Furthermore, the GVI-derived transformations increase the efficiency of the corresponding MCMC algorithms. Specifically, the normalized efficiencies are 0.0267 and 0.8762 for MALA-GVI and HMC-GVI, respectively, compared with 0.0002 and 0.8083 for the algorithms whose transformation matrices are estimated using preliminary MCMC samples.

\subsection{Efficient Bayesian Sampling Using the VAE-MH Algorithm}\label{Sec5.2}
We evaluate the sampling efficiency of the proposed VAE-MH algorithm on multimodal target distributions with different dimensions, tail behaviours, and numbers of modes. We compare VAE-MH with RMH and Diffusion-MH. Here, Diffusion-MH refers to the diffusion-model-assisted Metropolis--Hastings sampler adapted from \citet{J_CPC_hunt-smith2024accelerating}, in which a trained diffusion model is used to generate global proposal candidates and the Metropolis--Hastings correction is used to preserve the target distribution. Each sampler is run for $100{,}000$ sampling iterations. For RMH, the first $10{,}000$ iterations are discarded as burn-in. For VAE-MH and Diffusion-MH, the CrowdingDE-based initial training samples are prepended to the chain; these initial samples, together with the first $10{,}000$ sampling iterations, are removed before computing the reported metrics. Visual diagnostics are provided through corner plots for higher-dimensional examples and trajectory or sample plots for two-dimensional examples. Numerical performance is summarized by running time (Time), effective sample size per second (ESS/s), sampling efficiency (Efficiency), mode coverage (Mode cov.), mode frequency total variation distance (Mode TV), and empirical mode-transition rate (Mode trans.). Since a sampler can have a large ESS/s while remaining trapped in only part of a multimodal target, ESS/s is interpreted jointly with Mode cov., Mode TV, and Mode trans. All the reported values are averages over five independent runs.
\subsubsection{Sampling from mixtures of Gaussian and Student-$t$ distributions}
We first evaluate the proposed VAE-MH algorithm on two low-dimensional multimodal targets, following the experimental setting of \citet{C_ICLR_habib2019auxiliary}. Specifically, let
$
    \bm{\mu}_1=(10,0)^{\top},
    \bm{\mu}_2=(-10,0)^{\top},
    \bm{\Sigma}_1=\bm{\Sigma}_2=\mathbf I_2,
$
so that the Euclidean distance between the two component means is 20. The two-component Gaussian mixture target is defined as
$
    \pi_{\mathrm{G}}(\bm{\theta})
    =
    \omega\,
    \mathcal N(\bm{\theta}\mid \bm{\mu}_1,\bm{\Sigma}_1)
    +
    (1-\omega)\,
    \mathcal N(\bm{\theta}\mid \bm{\mu}_2,\bm{\Sigma}_2),
$
with $\omega=1/2$.

We also consider a heavier-tailed analogue based on Student $t$ components. Let $\nu=5$ denote the degrees of freedom. The Student $t$ mixture target is
$
    \pi_{\mathrm{T}}(\bm{\theta})
    =
    \omega\,
    t_{\nu}(\bm{\theta}\mid \bm{\mu}_1,\bm{\Sigma}_1)
    +
    (1-\omega)\,
    t_{\nu}(\bm{\theta}\mid \bm{\mu}_2,\bm{\Sigma}_2)
$
with $\omega=1/2$.

\begin{figure}[H]
    \centering
    \subfloat[Gaussian-2D\label{fig:2D_Gaussian_compare}]{
        \includegraphics[width=0.48\textwidth]{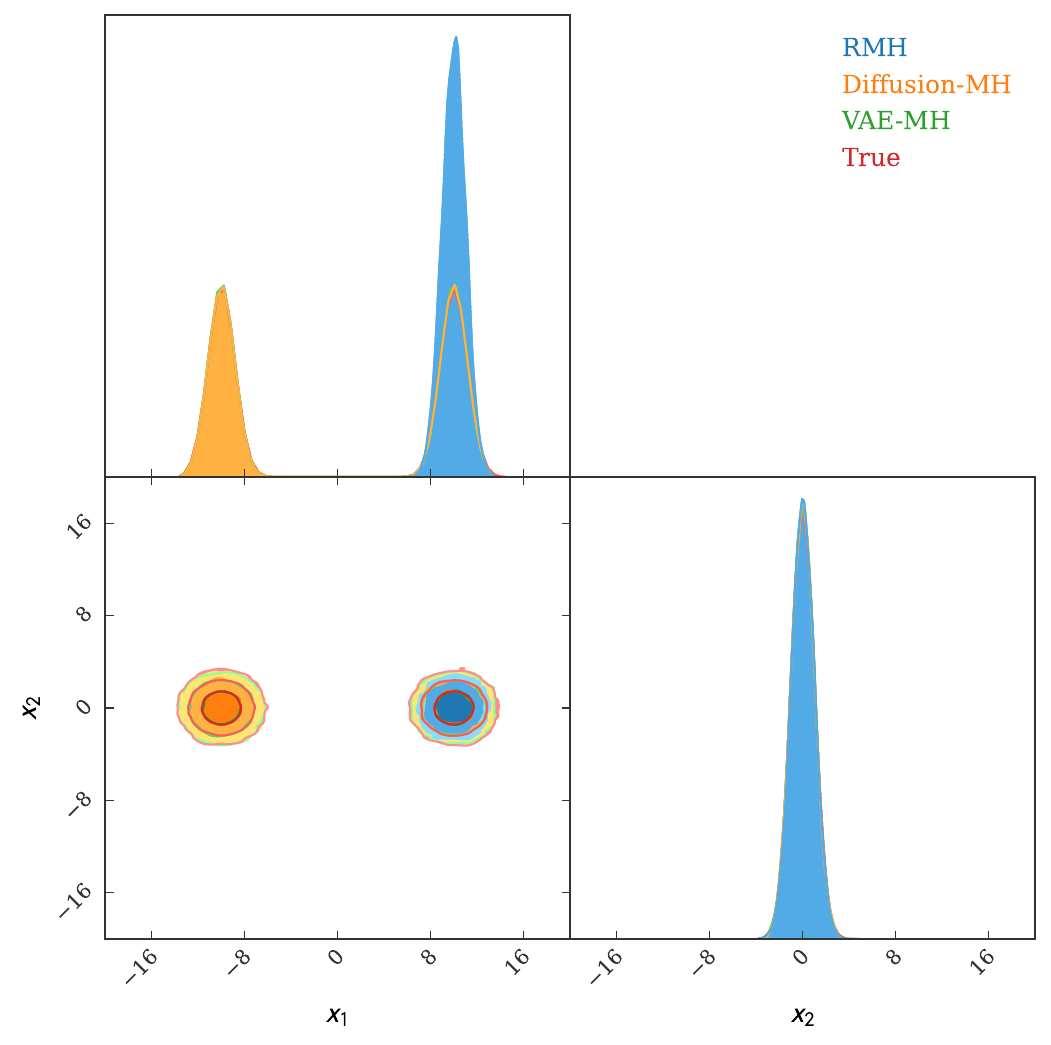}
    }
    \hfill
    \subfloat[Student-$t$\label{fig:2D_Student_compare}]{
        \includegraphics[width=0.48\textwidth]{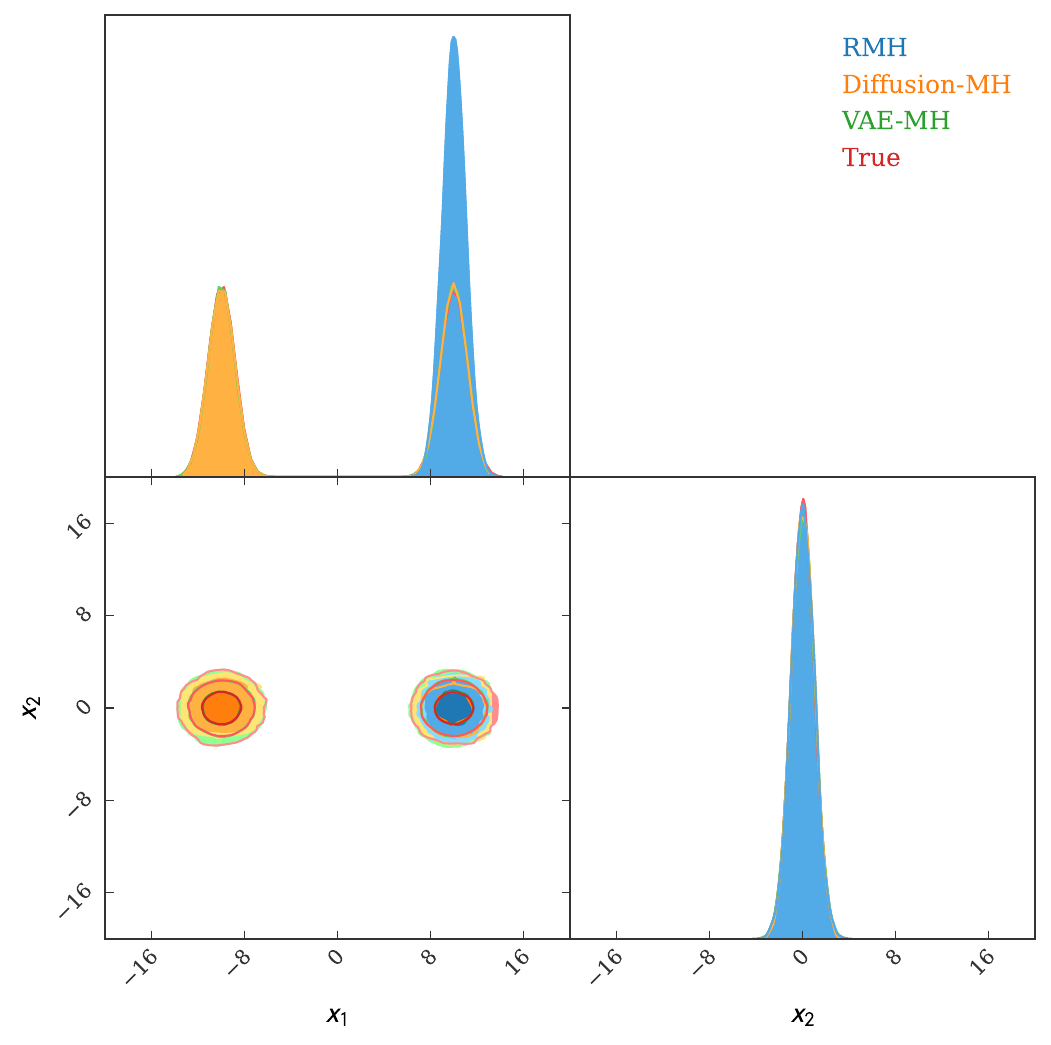}
    }

    \caption{Representative corner plots of the RMH, Diffusion-MH, and VAE-MH algorithms for sampling the two-dimensional Gaussian and Student-$t$ target distributions.}
    \label{fig:2D_corner_compare}
\end{figure}
\begin{table}[htbp]
    \centering
    \caption{Performance comparison of RMH, Diffusion-MH, and VAE-MH for sampling a 2-dimensional Gaussian mixture distribution. }
    \label{tab:gmm2}
    \small
    \setlength{\tabcolsep}{5pt}
    \begin{tabular}{lcccccc}
        \toprule
        Algorithm
         & Time (s)
         & ESS/s
         & E
         & Mode cov.
         & Mode TV
         & Mode trans.     \\
        \midrule
        RMH
         & \textbf{30.51}
         & 344.19
         & 0.1165
         & 1/2
         & 0.5000
         & 0.0000          \\

        Diffusion-MH
         & 95.15
         & \textbf{544.48}
         & 0.5755
         & \textbf{2/2}
         & \textbf{0.0016}
         & 0.3646          \\

        VAE-MH
         & 131.67
         & 486.22
         & \textbf{0.7105}
         & \textbf{2/2}
         & \textbf{0.0016}
         & \textbf{0.3960} \\
        \bottomrule
    \end{tabular}
\end{table}
\begin{table}[htbp]
    \centering
    \caption{Performance comparison of RMH, Diffusion-MH, and VAE-MH for sampling a 2-dimensional Student-$t$ mixture distribution. Reported values are averages over five independent runs.}
    \label{tab:student2}
    \small
    \setlength{\tabcolsep}{5pt}
    \begin{tabular}{lcccccc}
        \toprule
        Algorithm
         & Time (s)
         & ESS/s
         & E
         & Mode cov.
         & Mode TV
         & Mode trans.     \\
        \midrule
        RMH
         & \textbf{14.65}
         & 381.55
         & 0.0620
         & 1.4/2
         & 0.3791
         & 0.0000          \\

        Diffusion-MH
         & 97.86
         & \textbf{385.31}
         & \textbf{0.4186}
         & \textbf{2/2}
         & 0.0024
         & \textbf{0.3567} \\

        VAE-MH
         & 165.83
         & 214.51
         & 0.3948
         & \textbf{2/2}
         & \textbf{0.0014}
         & 0.3239          \\
        \bottomrule
    \end{tabular}
\end{table}
Figure~\ref{fig:2D_corner_compare} and Tables~\ref{tab:gmm2}--\ref{tab:student2} summarize the performance of the three samplers on the two-dimensional Gaussian and Student $t$ mixture targets. The corner plots show that RMH has difficulty moving between the two separated modes. In the Gaussian mixture case, it remains in only one component, with zero mode-transition rate and a Mode TV distance of 0.5000. In the heavier-tailed Student $t$ mixture, RMH occasionally reaches both modes across repeated runs, but its average mode-transition rate is still essentially zero, indicating poor cross-mode mixing.

Both Diffusion-MH and VAE-MH substantially improve global exploration. They cover both modes in both experiments and achieve very small Mode TV distances, showing that the empirical mode frequencies are close to the true mixture weights. Diffusion-MH attains the highest ESS/s in both examples, reflecting faster sampling per unit time. In the Gaussian mixture case, VAE-MH achieves the highest sampling efficiency, while both generative-proposal methods have similarly small Mode TV distances. For the heavier-tailed Student $t$ mixture, Diffusion-MH has slightly larger sampling efficiency and mode-transition rate, whereas VAE-MH gives a comparable mode coverage and a smaller Mode TV distance. Overall, the results confirm that global generative proposals can mitigate the mode-trapping behavior of random-walk MH, while the relative advantage between Diffusion-MH and VAE-MH depends on the target distribution and the performance metric being considered.
\subsubsection{Sampling from a 10-dimensional Gaussian mixture}

We next consider a two-component Gaussian mixture in ten dimensions, following the setting of \citet{J_CPC_hunt-smith2024accelerating}. The target density is
\begin{equation*}
    \pi_{10}(\bm{\theta})
    =
    w_A \frac{\exp\{-\|\bm{\theta}-\bm{\theta}_A\|^2/2\}}{(2\pi)^{d/2}}
    +
    w_B \frac{\exp\{-\|\bm{\theta}-\bm{\theta}_B\|^2/2\}}{(2\pi)^{d/2}},
\end{equation*}
where $d=10$, $w_A=2/3$, $w_B=1/3$,
$\bm{\theta}_A=(8,3,\mathbf{0}_8^\top)^\top$, and
$\bm{\theta}_B=(-2,3,\mathbf{0}_8^\top)^\top$.
\begin{table}[htbp]
    \centering
    \caption{Performance comparison of RMH, Diffusion-MH, and VAE-MH for sampling a 10-dimensional Gaussian mixture distribution. Reported values are averages over five independent runs.}
    \label{tab:gmm10}
    \small
    \setlength{\tabcolsep}{5pt}
    \begin{tabular}{lcccccc}
        \toprule
        Algorithm
         & Time (s)
         & ESS/s
         & E
         & Mode cov.
         & Mode TV
         & Mode trans      \\
        \midrule
        RMH
         & \textbf{32.18}
         & 69.59
         & 0.0249
         & 1/2
         & 0.6000
         & 0.0000          \\

        Diffusion-MH
         & 419.53
         & 74.81
         & \textbf{0.3484}
         & \textbf{2/2}
         & 0.0035
         & \textbf{0.2684} \\

        VAE-MH
         & 220.23
         & \textbf{130.79}
         & 0.3129
         & \textbf{2/2}
         & \textbf{0.0024}
         & 0.2225          \\
        \bottomrule
    \end{tabular}
\end{table}
\begin{figure}[H]
    \centering
    \includegraphics[width=1\textwidth]{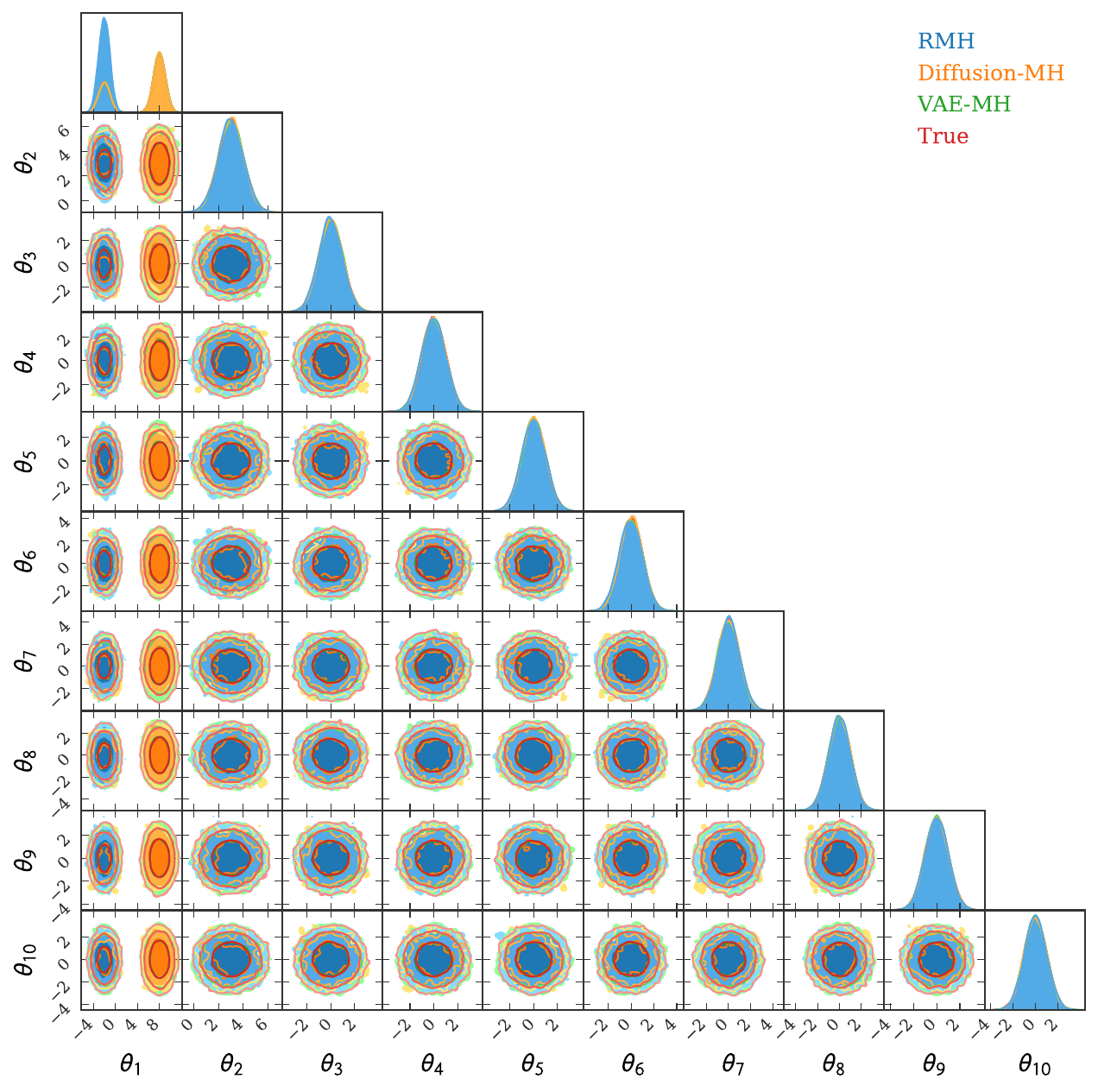}
    \caption{Representative corner plot of the RMH, Diffusion-MH, and VAE-MH algorithms for sampling the 10-dimensional Gaussian mixture distribution.}
    \label{fig:10D_GAUSSIAN_compare}
\end{figure}
Figure~\ref{fig:10D_GAUSSIAN_compare} and Table~\ref{tab:gmm10} summarize the performance of the three samplers on the 10-dimensional Gaussian mixture. The corner plot shows that RMH samples only one component: the marginal distribution of $\theta_1$ is unimodal, and the pairwise plots involving $\theta_1$ do not show the second cluster. Both Diffusion-MH and VAE-MH recover the two components of the target distribution. Their contours in the corner plot are close to the true contours, and both methods achieve full mode coverage with small mode-TV distance. The main difference is computational. Diffusion-MH has a slightly higher efficiency and mode-transition rate, but it requires substantially longer runtime. VAE-MH attains a higher ESS per second while using less time, and its corner plot remains close to the target across the pairwise projections.

\subsubsection{Sampling from 2-dimensional Gaussian mixture of 20 components}
Following the experimental setup of \citet{J_JASA_liang2001realparameter}, the target distribution is defined as:
\begin{equation*}
    \pi_{20}(\bm{\theta})
    =
    \sum_{i=1}^{20}
    w_i
    \mathcal N(\bm{\theta}\mid \bm{\mu}_i,\sigma^2\mathbf I_2)
    =
    \sum_{i=1}^{20}
    w_i
    \frac{1}{2\pi\sigma^2}
    \exp\left\{
    -\frac{1}{2\sigma^2}
    (\bm{\theta}-\bm{\mu}_i)^{\top}
    (\bm{\theta}-\bm{\mu}_i)
    \right\},
\end{equation*}
where $\sigma=0.1$, $w_1=\cdots=w_{20}=0.05$, and the mean vectors $\bm{\mu}_1,\bm{\mu}_2,\cdots,\bm{\mu}_{20}$ are uniformly drawn from the rectangle $[0,10] \times [0,10]$, as listed in Appendix Table~\ref{tab:mean_gmm20}.
\begin{figure}[ht]
    \centering
    \includegraphics[width=0.5\linewidth]{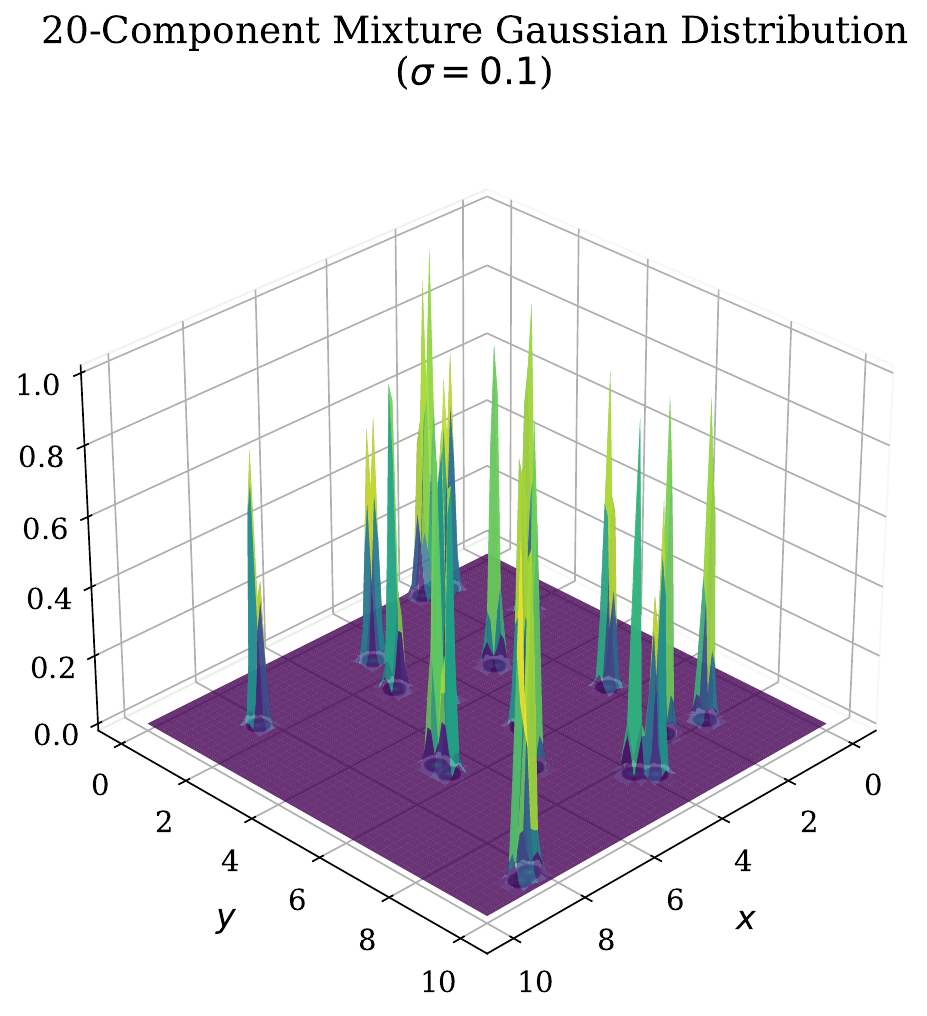}
    \caption{Density surface of the 20-component two-dimensional Gaussian
        mixture with equal component weights and $\sigma=0.1$. The component means
        are located in $[0,10]^2$, producing narrow isolated peaks separated by
        regions with negligible probability mass.}
    \label{fig:gmm20_density}
\end{figure}

\begin{figure}[H]
    \centering
    \includegraphics[width=1.0 \linewidth]{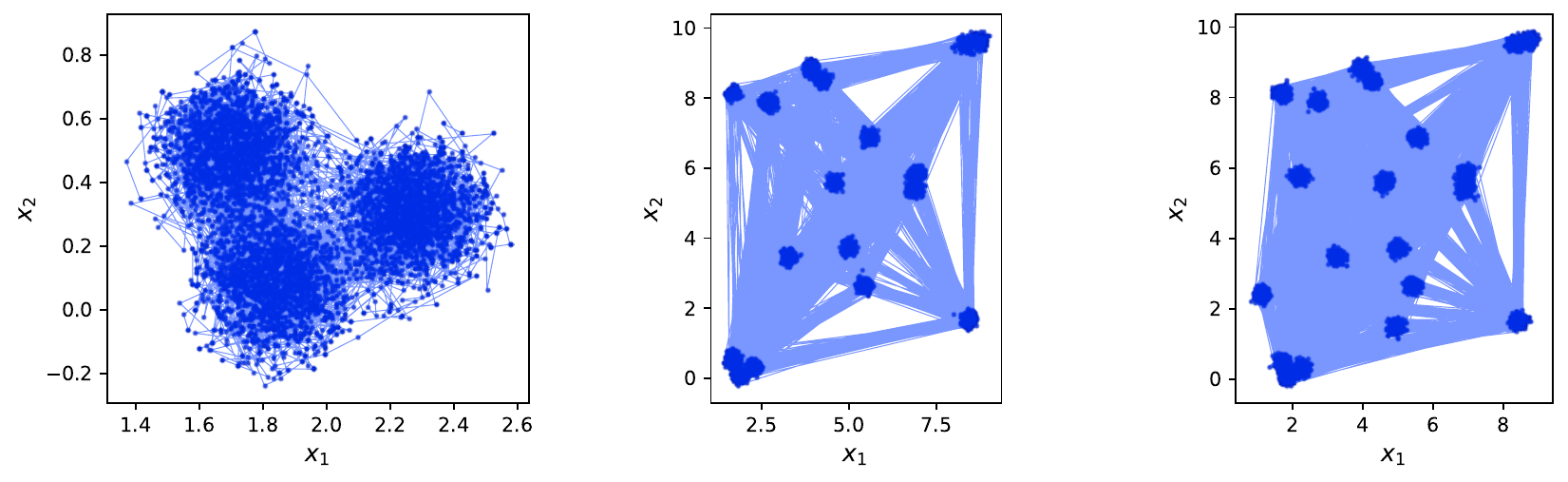}
    \caption{Representative trajectory plot of the first 10000 post-burn-in samples of random-walk MH, Diffusion-MH and VAE-MH algorithms.}
    \label{fig:trace_plot}
\end{figure}

\begin{figure}[H]
    \centering
    \includegraphics[width=1.0 \linewidth]{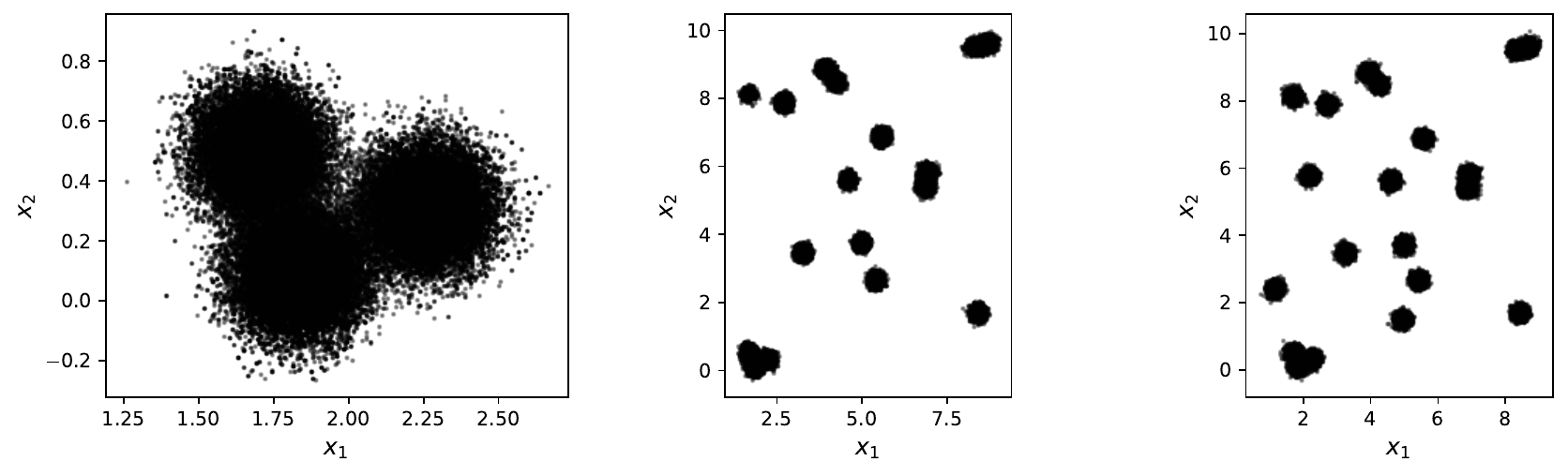}
    \caption{Representative sample plot of the post-burn-in random-walk MH, Diffusion-MH and VAE-MH chains.}
    \label{fig:cluster}
\end{figure}
\begin{table}[htbp]
    \centering
    \caption{Performance comparison of RMH, Diffusion-MH, and VAE-MH for sampling a 2-dimensional Gaussian mixture distribution with 20 components. Reported values are averages over five independent runs.}
    \label{tab:gmm20}
    \small
    \setlength{\tabcolsep}{5pt}
    \begin{tabular}{lcccccc}
        \toprule
        Algorithm
         & Time (s)
         & ESS/s
         & E
         & Mode cov.
         & Mode TV
         & Mode trans.     \\
        \midrule
        RMH
         & \textbf{28.99}
         & 74.26
         & 0.0239
         & 3/20
         & 0.8500
         & 0.0532          \\

        Diffusion-MH
         & 97.90
         & 191.01
         & 0.2078
         & 17.4/20
         & 0.2060
         & 0.3486          \\

        VAE-MH
         & 275.04
         & \textbf{214.91}
         & \textbf{0.6568}
         & \textbf{20/20}
         & \textbf{0.0077}
         & \textbf{0.7597} \\
        \bottomrule
    \end{tabular}
\end{table}
Figures~\ref{fig:trace_plot} and~\ref{fig:cluster} compare the mixing behavior of the three samplers. The RMH sampler stays in a local region for a long period and does not move effectively between separated components.  Both Diffusion-MH and VAE-MH improve cross-mode exploration: Diffusion-MH visits 17.4 out of the 20 components on average, whereas VAE-MH recovers all 20 components. The trajectory plot also shows better and more stable transitions for VAE-MH, indicating better mixing across the separated modes. The numerical results in Table~\ref{tab:gmm20} are consistent with these visual findings. RMH has the lowest runtime, but its low efficiency and poor mode coverage show that it is not sampling the full target distribution adequately. Diffusion-MH improves global movement, but it does not consistently cover all components and has a larger mode-frequency error. VAE-MH achieves best overall performance with full mode coverage, a smaller mode-frequency error and better mixing behavior.

\section{Discussion}\label{sec6}
The numerical studies suggest that VI can improve MCMC most effectively when it is used as a source of structural information rather than as a direct replacement for sampling. In the HMC experiments, Gaussian variational approximations provide fast estimates of the target covariance structure, which can be used to construct fixed linear transformations before sampling. This is especially useful when covariance estimation from preliminary MCMC runs is expensive, as seen in the high-dimensional Gaussian and GLMM examples. The results indicate that the benefit of GVI is not only computational speed, but also a stable way to obtain geometry information for gradient-based samplers.

For multimodal targets, the VAE-MH experiments show a different role of VI. Instead of approximating the target density for inference directly, the VAE-type model is used to generate global proposals that help the Markov chain move between separated modes. The local random-walk proposal then explores within each mode, while the MH correction reduces the bias introduced by the learned proposal. The experiments on Gaussian and Student-$t$ mixtures show that this global-local strategy can substantially improve mode coverage and cross-mode movement compared with random-walk MH. These results support the view that generative models can be useful proposal mechanisms, particularly when they are combined with mode finding and an explicit accept-reject correction.

There are also several limitations. The GVI-based transformation uses a fixed global covariance estimate, so it may be less effective for strongly non-Gaussian or highly multimodal targets where a single linear transformation cannot represent local geometry well. The VAE-MH sampler depends on the quality of the mode-finding stage and on the ability of the VAE proposal to learn from the initial and adaptively collected samples. In addition, the KDE-based proposal-density correction used in VAE-MH may become challenging in substantially higher dimensions. Future work will investigate more scalable proposal-density estimators, stronger theoretical guarantees for the adaptive global-local sampler, and applications to more realistic Bayesian models with complex multimodal posterior structures. We will also explore combining other generative models with advanced MCMC kernels, including reversible jump MCMC \citep{J_BIOMET_green1995reversible}, for problems involving parameter spaces of varying dimensions.



\bibliographystyle{apacite}
\bibliography{bibs/abbr}

\appendix

\section{Algorithmic details}
\label{app:algorithmic-details}
The algorithms referred to in Sections~\ref{sec:hmc-gvi} and~\ref{sec3.2} are collected here.
\begin{algorithm}[H]
    \caption{GVI-preconditioned Hamiltonian Monte Carlo}
    \label{alg:hmc-gvi}
    \begin{algorithmic}[1]
        \Require Log unnormalized target density $\ell(\bm{\theta})=\log\tilde{\pi}(\bm{\theta})$, gradient $\nabla_{\bm{\theta}}\ell(\bm{\theta})$, GVI routine, step size $\epsilon$, number of leapfrog steps $L$, burn-in size $B$, and number of retained samples $N$
        \Ensure Samples $\bm{\theta}^{(1)},\ldots,\bm{\theta}^{(N)}$
        \State Obtain
        $q_{\mathrm{GVI}}(\bm{\theta})
            =
            \mathcal N
            (
            \bm{\mu}_{\mathrm{GVI}},
            \bm{\Sigma}_{\mathrm{GVI}}
            )$
        by GVI
        \State Set
        $\mathbf S\gets \bm{\Sigma}_{\mathrm{GVI}}$
        and initialize
        $\bm{\theta}\gets \bm{\mu}_{\mathrm{GVI}}$
        \State Compute
        $\ell\gets \ell(\bm{\theta})
            =
            \log \tilde{\pi}(\bm{\theta})$
        \State Compute a lower-triangular Cholesky factor
        $\mathbf C=\operatorname{chol}(\mathbf S)$,
        so that
        $\mathbf C\mathbf C^{\top}=\mathbf S$

        \For{$t=1$ to $B+N$}
        \State Draw $\mathbf z\sim \mathcal N(\mathbf 0,\mathbf I_d)$ and compute $\bm{\psi}$ by solving $\mathbf C^{\top}\bm{\psi}=\mathbf z$
        \Statex \Comment{Then $\bm{\psi}=\mathbf C^{-\top}\mathbf z\sim\mathcal N(\mathbf 0,\mathbf S^{-1})$}

        \State Compute
        $
            K
            \gets
            \frac12
            \bm{\psi}^{\top}
            \mathbf S
            \bm{\psi}
        $

        \State Set
        $\bm{\theta}^{\star}\gets \bm{\theta}$
        and
        $\bm{\psi}^{\star}\gets \bm{\psi}$

        \State
        \[
            \bm{\psi}^{\star}
            \gets
            \bm{\psi}^{\star}
            +
            \frac{\epsilon}{2}
            \nabla_{\bm{\theta}}
            \ell(\bm{\theta}^{\star})
        \]

        \For{$l=1$ to $L$}
        \State
        \[
            \bm{\theta}^{\star}
            \gets
            \bm{\theta}^{\star}
            +
            \epsilon
            \mathbf S
            \bm{\psi}^{\star}
        \]
        \If{$l\neq L$}
        \State
        \[
            \bm{\psi}^{\star}
            \gets
            \bm{\psi}^{\star}
            +
            \epsilon
            \nabla_{\bm{\theta}}
            \ell(\bm{\theta}^{\star})
        \]
        \EndIf
        \EndFor

        \State
        \[
            \bm{\psi}^{\star}
            \gets
            \bm{\psi}^{\star}
            +
            \frac{\epsilon}{2}
            \nabla_{\bm{\theta}}
            \ell(\bm{\theta}^{\star})
        \]

        \State
        $\bm{\psi}^{\star}
            \gets
            -\bm{\psi}^{\star}$

        \State Compute
        $
            \ell^{\star}
            \gets
            \ell(\bm{\theta}^{\star})
            =
            \log \tilde{\pi}(\bm{\theta}^{\star})
        $
        and
        $
            K^{\star}
            \gets
            \frac12
            (\bm{\psi}^{\star})^{\top}
            \mathbf S
            \bm{\psi}^{\star}.
        $

        \State Compute
        $
            \log a
            \gets
            \ell^{\star}
            -
            \ell
            +
            K
            -
            K^{\star}.
        $

        \State Draw $u\sim \mathrm{Uniform}(0,1)$

        \If{$\log u < \min\{0,\log a\}$}
        \State
        $\bm{\theta}
            \gets
            \bm{\theta}^{\star}$
        \State
        $\ell\gets \ell^{\star}$
        \EndIf

        \If{$t>B$}
        \State Store
        $\bm{\theta}^{(t-B)}
            \gets
            \bm{\theta}$
        \EndIf
        \EndFor
    \end{algorithmic}
\end{algorithm}

\begin{algorithm}[H]
    \caption{Crowding-based Differential Evolution for mode finding}%
    \label{alg:differential_evolution}
    \begin{algorithmic}[1]
        \Require Objective $f(\bm{x})=\tilde{\pi}(\bm{x})$ to be maximized, search domain $\Omega$, $N$, $F$, $CR$, $E_{\max}$, and number of modes $J$
        \Ensure Estimated mode locations $\widehat{\bm{m}}_1,\ldots,\widehat{\bm{m}}_J$
        \State Randomly initialize current population $P=\{\bm{x}_1,\ldots,\bm{x}_{N}\}\subset\Omega$
        \While{the number of target evaluations is less than $E_{\max}$}
        \For{$i=1,\ldots,N$}
        \State Select distinct indices $r_1,r_2,r_3\in\{1,\ldots,N\}\setminus\{i\}$
        \State $\bm{v}_{i}\gets \bm{x}_{r_1}+F(\bm{x}_{r_2}-\bm{x}_{r_3})$
        \State Draw $j_{\mathrm{rand}}\sim U\{1,\ldots,d\}$
        \For{$j=1,\ldots,d$}
        \If{$\operatorname{rand}()\leq CR$ or $j=j_{\mathrm{rand}}$}
        \State $u_{i,j}\gets v_{i,j}$
        \Else
        \State $u_{i,j}\gets x_{i,j}$
        \EndIf
        \EndFor
        \State $k_i\gets \arg\min_{k=1,\ldots,N}\|\bm{x}_{k}-\bm{u}_{i}\|$
        \If{$f(\bm{u}_{i})\geq f(\bm{x}_{k_i})$}
        \State Replace $\bm{x}_{k_i}$ by $\bm{u}_{i}$ in $P$
        \EndIf
        \EndFor
        \EndWhile
        \State Cluster the final population into $J$ groups
        \State In each cluster, choose the individual with the largest value of $f$ as $\widehat{\bm{m}}_j$
    \end{algorithmic}
\end{algorithm}

\begin{algorithm}[H]
    \caption{VAE--MH sampler for multi-modal targets}
    \label{VAE-MH}
    \begin{algorithmic}[1]
        \Require Target density $\pi$ up to a normalizing constant, CrowdingDE mode-finding routine, $n_0$, $\tau$, $W$, $M$, $h$, $\rho$, $\sigma_{\mathrm{rw}}$, and number of refresh rounds $R$
        \Ensure Chain output $\mathcal C=\{\bm{\theta}^{(1)},\ldots,\bm{\theta}^{(R\tau)}\}$
        \State Obtain approximate modes $\widehat{\bm{m}}_1,\ldots,\widehat{\bm{m}}_J$ using CrowdingDE
        \State Construct $\mathcal{D}_0$ from local Gaussian samples around $\widehat{\bm{m}}_1,\ldots,\widehat{\bm{m}}_J$
        \State Initialize the chain at a random point in $\mathcal{D}_0$
        \State Set initial proposal pool $\mathcal P_0\gets \mathcal D_0$
        \State Set training history $\mathcal H\gets \mathcal D_0$ and chain output $\mathcal C\gets\emptyset$
        \For{$r=0,1,\ldots,R-1$}
        \State Fit the KDE proposal density $q_r$ to $\mathcal P_r$ as in equation~\eqref{eq:kde_proposal}
        \For{$s=1,\ldots,\tau$}
        \State Draw $u\sim U(0,1)$
        \If{$u<\rho$}
        \State Propose $\bm{\theta}'$ from the KDE proposal $q_r$
        \State Accept $\bm{\theta}'$ with probability in equation~\eqref{eq:vae_mh_acceptance}
        \Else
        \State Propose $\bm{\theta}'=\bm{\theta}^{(t)}+\bm{\epsilon}$, $\bm{\epsilon}\sim\mathcal{N}(\bm{0},\sigma_{\mathrm{rw}}^2\mathbf{I}_d)$
        \State Accept $\bm{\theta}'$ with random-walk MH probability
        \EndIf
        \State Append the current state to $\mathcal C$ and $\mathcal H$
        \EndFor
        \State Set $\mathcal{D}_{r+1}$ to the most recent $W$ elements of $\mathcal H$, or all available elements if fewer than $W$
        \State Retrain a VAE-type generative model $G_{r+1}$ on $\mathcal{D}_{r+1}$
        \State Generate $\mathcal P_{r+1}=\{\bm{\zeta}_{r+1,1},\ldots,\bm{\zeta}_{r+1,M}\}$ from $G_{r+1}$
        \EndFor
    \end{algorithmic}
\end{algorithm}

\section{Additional experiment details}
\label{app:additional-experiment-details}
The mean vectors used in the 20-component Gaussian mixture experiment are listed in Table~\ref{tab:mean_gmm20}.
\begin{table}[htbp]
    \centering
    \caption{Mean vectors of the 20 components of the mixture Gaussian distribution}
    \label{tab:mean_gmm20}
    \begin{tabular}{*{4}{c S[table-format=1.2] S[table-format=1.2]}}
        \toprule
        $i$ & {$\mu_{i1}$} & {$\mu_{i2}$} &
        $i$ & {$\mu_{i1}$} & {$\mu_{i2}$} &
        $i$ & {$\mu_{i1}$} & {$\mu_{i2}$} &
        $i$ & {$\mu_{i1}$} & {$\mu_{i2}$}                                                          \\
        \midrule
        1   & 2.18         & 5.76         & 6  & 3.25 & 3.47 & 11 & 5.41 & 2.65 & 16 & 4.93 & 1.50 \\
        2   & 8.67         & 9.59         & 7  & 1.70 & 0.50 & 12 & 2.70 & 7.88 & 17 & 1.83 & 0.09 \\
        3   & 4.24         & 8.48         & 8  & 4.59 & 5.60 & 13 & 4.98 & 3.70 & 18 & 2.26 & 0.31 \\
        4   & 8.41         & 1.68         & 9  & 6.91 & 5.81 & 14 & 1.14 & 2.39 & 19 & 5.54 & 6.86 \\
        5   & 3.93         & 8.82         & 10 & 6.87 & 5.40 & 15 & 8.33 & 9.50 & 20 & 1.69 & 8.11 \\
        \bottomrule
    \end{tabular}
\end{table}
\end{document}